\documentclass[manuscript,12pt,anonymous=false,nonacm]{acmart}
\AtBeginDocument{%
  \providecommand\BibTeX{{%
    \normalfont B\kern-0.5em{\scshape i\kern-0.25em b}\kern-0.8em\TeX}}}

\setcopyright{acmcopyright}
\copyrightyear{2020}
\acmYear{2020}
\acmDOI{10.1145/1122445.1122456}

\acmConference[FAccT '21]{FAccT '21: ACM Conference on Fairness, Accountability, and Transparency}{March 2021}{Virtual}
\acmBooktitle{FAccT '21: ACM Conference on Fairness, Accountability, and Transparency,
  March 2021, Virtual}
\acmPrice{15.00}
\acmISBN{978-1-4503-XXXX-X/18/06}

\usepackage[font=small,labelfont=sc]{caption}
\usepackage{subcaption}
\usepackage{url}

\usepackage{enumitem}
\setlist[enumerate]{topsep=0pt,itemsep=-1ex,partopsep=1ex,parsep=1ex}

\setcopyright{none} 
\renewcommand\footnotetextcopyrightpermission[1]{} 

\usepackage{float}

\setcitestyle{round}
\usepackage{todonotes}
\usepackage{mathtools}

\begin{document}

\title{Navigating Epistemic Monocultures in AI-Driven Science: A Simulation Study}
\author{Sina Fazelpour}
\affiliation{%
   \institution{Northeastern University}
   \city{Boston}
    \country{United States}
}
 \email{s.fazel-pour@northeastern.edu}
\author{Joseph O'Brien}
\affiliation{%
   \institution{University of California, San Diego}
   \city{San Diego}
    \country{United States}
}
\email{j3obrien@ucsd.edu}
\author{Hannah Rubin}
\affiliation{%
   \institution{University of Missouri}
   \city{Columbia}
    \country{United States}
}
\email{hmrfyb@missouri.edu}

\begin{abstract}
AI integration into scientific communities promises accelerated discovery but raises concerns about detrimental homogenization. We develop an NK landscape model to explore these promises and risks. We find that non-personalized AI systems that offer uniform guidance yield benefits only under a narrow conjunction of problem structure, practices, and baseline research capabilities, becoming harmful otherwise. We implement two proposed mitigations: randomization and personalization. While randomization's utility remains restricted to decomposable problems, personalization can enhance diversity, enabling benefits across a broader range of conditions. Crucially, these benefits are not automatic, but depend on effective institutional adaptation, requiring new standards and practices.
\end{abstract}

\maketitle
\pagestyle{plain}
\sloppy

\section{Introduction}\label{sec:intro}
Artificial intelligence (AI) tools are increasingly integrated across the scientific pipeline, from hypothesis generation and data analysis to simulating research participants and automating laboratory procedures. This integration raises both hopes and concerns. On the one hand, for some researchers, AI integration promises to accelerate discovery by allowing diverse teams to perform complex inquiries, more efficiently and less hampered by resource or expertise disparities~\citep{hume2025ai,wang2023scientific,gottweis2025towards}. On the other hand, it has been argued that the widespread adoption of AI may lead to the formation of epistemic monocultures, reducing the productive heterogeneity of methodologies with which scientists pursue their work~\citep{messeri2024artificial,burton2024large}.

In this paper, we develop a simulation model to explore when the adoption of AI tools in scientific communities may result in such detrimental homogenization, and what strategies might mitigate this risk. Specifically, we use the NK landscape framework~\citep{kauffman1987towards} to model scientific problem-solving as a search through complex solution spaces, and examine how different AI designs affect epistemic success and diversity across different problem structures.

Our results reveal that the impacts of AI adoption depend critically on AI system design, problem structure, and institutional norms and behaviors in the relevant communities. We first look at a type of \textit{non-personalized} AI systems, which recommends uniform best practices without accounting for problem-solving context of querying epistemic agents. We find that this yields benefits only under highly specific conditions: when problems are easily decomposed into smaller sub-problems, such that those aspects handled by AI can be tackled somewhat independently, and when scientists rely on AI only moderately for their research. Outside these conditions, AI adoption results in premature convergence toward epistemic monocultures and harms longer term collective performance. 

We implement two mitigation strategies inspired by proposals in the literature for counteracting the risks of AI-driven homogenization~\citep{fugener2021will,jain2024position}: \textit{randomization}, which introduces diversity in the selection of good recommendations by sampling from top-performing solutions rather than recommending a single best one; and \textit{personalization}, which tailors recommendations to each agent's specific context. We find that, while useful in some settings, Randomization does not reliably increase performance across different problem-solving contexts; its utility remains restricted to the same structural conditions as Non-personalized AI. In contrast, Personalization proves robustly beneficial, albeit to varying extents, across various use rates and different kinds of problems epistemic communities may be attempting to solve.\footnote{Terminological note: we use capitalization to distinguish between Randomization and Personalization \textit{as implemented in our model} from the broader real-world phenomena, in order to create some separation between our findings and potential policy implications, which would require additional assumptions (e.g., see Section~\ref{sec:discussion}).}

Importantly, our findings suggest that productive AI integration requires more than tool adoption. For Non-personalized AI systems, the requisite problem structure is often an achievement of organizational practices, depending on standardization, established protocols, and divisions of labor, rather than simply an inherent feature of scientific domains. For Personalized AI systems, effectiveness requires established institutional standards for documenting and communicating often tacit human and organizational factors that shape scientific practice, and can vary based on how communities adapt their exploratory practices to complement AI capabilities. We discuss these implications for scientific communities navigating AI adoption, along with limitations of our approach and directions for future research.

The rest of the paper is structured as follows. In \S\ref{sec:background}, we review previous work on AI in science, concerns about monoculture formation, and the NK landscape framework as an appropriate tool for examining these dynamics. \S\ref{sec:methods} provides the formal details of our model implementation. In \S\ref{sec:simulation1-non-personalized} and \S\ref{sec:simulation2-mitigation}, we present simulation results exploring monoculture formation and the effectiveness of various interventions. Finally, \S\ref{sec:discussion} discusses implications for scientific practice. 
\section{Background and Related Work}\label{sec:background}
\subsection{AI in Scientific Communities}
AI systems are integrated into every stage of the research pipeline. They are employed to synthesize vast bodies of literature, formulate novel hypotheses, automate experimental designs, generate synthetic data, and perform multiverse analysis~\citep{wang2023scientific,zhang2025exploring,lu2024unleashing,gottweis2025towards,si2024can,gao2025take,messeri2024artificial,bertran2026many}.\footnote{\citet{messeri2024artificial} provide a useful taxonomy of these applications across the research pipeline.} These varied use cases draw on a correspondingly broad set of technical methods and architectures for tasks such as literature synthesis, prediction, data generation, and the optimization of experimental protocols. More recently, agentic and coding-capable AI systems have been used to orchestrate these diverse capabilities into multi-agent workflows that automate ever larger portions of the scientific pipeline with minimal human intervention~\citep{baek2025researchagent,schmidgall2025agent,lu2026towards}.

Across this variety in design and application, advocates highlight the shared promise of AI-driven science to accelerate scientific discovery. Beyond automating and speeding up routine tasks, AI systems are seen as broadening both the type of tasks that can be performed and the community that can perform them. For example, AI tools are seen as offering the promise of more efficiently navigating complex problem spaces, by identifying promising, but neglected directions of inquiry or novel patterns across vast, disparate datasets~\citep{wang2023scientific,sourati2023accelerating}. Moreover, these systems are said to ``democratize'' science by facilitating access to capabilities that previously required deep domain expertise and specialized resources~\citep{dessimoz2024ai,hume2025ai}. Similarly, emerging ``cloud laboratories'' and AI-driven experimental platforms may allow researchers to remotely execute and reproduce experiments, making procedures accessible to and implementable by a broader community~\citep{lu2024unleashing,adam2024automated}. 

\subsection{Epistemic Monocultures}
Despite these promises, a growing body of literature warns that the widespread reliance on algorithmic tools may inadvertently reduce the diversity of scientific inquiry in particular, and that of epistemic communities more broadly~\citep{messeri2024artificial, kleinberg2021algorithmic,burton2024large,hao2026artificial}. In particular, \citet{messeri2024artificial} distinguish between two types of such \textit{epistemic monocultures}: \textit{monocultures of knowers} (homogeneity in standpoints from which science is done) and \textit{monocultures of knowing} (homogeneity in how science is done). With respect to the latter, which will be our focus in this paper, they argue that the widespread adoption of AI tools risks producing a monoculture by prioritizing quantitative ways of knowing that appear portable and scalable, but which systematically ``strip out the contextual sensitivity and local details'' required for deep understanding~\citep*[][p. 54]{messeri2024artificial}.

On Messeri and Crockett's account, a key driver of this homogenization is the way in which the technical success of AI tools can obscure their underlying methodological and theoretical commitments. As Messeri and Crockett observe, the construction of any computational tool requires developers to make various choices that inevitably embed specific assumptions and values into the resulting artifacts. When resource- or expertise-constrained teams adopt these ``high-impact'' artifacts to study similar questions, however, they may do so without appreciating the crucial import of those hidden, context-specific choices. This can lead scientific communities to fall prey to an ``illusion of exploratory breadth''~\citep{messeri2024artificial}: researchers believe they are exploring the full landscape of potential solutions, when in reality they are converging on a narrow subset of hypotheses that are legible to the dominant algorithmic paradigm.

Importantly, Messeri and Crockett's concern is not restricted to cases in which researchers consciously defer to particular AI systems on discrete research questions. As these systems become part of standard scientific infrastructure, they can embed consequential choices---e.g., how a research question is operationalized, which sources are consulted, or how results are filtered and ranked---that users may adopt without insights into what those choices are, how they were made, and why. In this way, AI systems can inconspicuously curate the space of methodological options that researchers encounter, adopt, and build upon, without appearing to restrict those choices at all~\citep[see also][]{passi2025agentic}.

\subsection{Modeling the Impact of AI Adoption in Scientific Communities}
How should we evaluate the systemic impacts of AI integration in scientific communities in light of these promises and risks? To address this, we employ the NK landscape framework as a generalized model of scientific problem-solving~\citep{kauffman1987towards}. NK landscapes have been fruitfully employed in philosophical and social scientific research to examine many of the aforementioned factors that arise in the context of AI integration in science: the complexity of a scientific problem; potential interdependence between different aspects of a problem (e.g., those amenable to quantitative methods and those that are not); relative (in)efficiencies of different methodologies; the promises and failure modes of communication; and the value of diversity and the risks of homogenization in scientific communities~\citep{lazer2007network,gomez2019clustering,wu2024better,wu2023should,muldoon2013diversity,reijula2023division,huang2024landscapes,grim2013scientific}. Appropriately augmented, we suggest, this framework can offer a productive lens for understanding the systemic effects of widespread AI adoption on scientific search at a useful level of abstraction. 

Importantly, the abstraction afforded by the framework allows us to draw more generalizable lessons than implementation-specific approaches, which can be hampered by the diversity of AI applications and architectures. At the same time, it allows us to move beyond general concerns about monocultures to explore specific trade-offs, and the conditions under which the benefits of AI adoption could plausibly outweigh its risks. In the following section, we formally describe the model and how we extend it to capture issues surrounding AI integration in science. In Section~\ref{sec:simulation1-non-personalized}, we return to the dynamics of monoculture formation described by Messeri and Crockett.  
\section{General Modeling Framework}\label{sec:methods}
In this section, we describe the NK framework and how we extend it to model AI integration in scientific communities. To make the formalisms concrete, we employ a running example of drug discovery in public health contexts. 

\subsection{The $NK$ Landscape and Problem Complexity}
Consider research teams developing treatments for a neglected disease. Such teams face a range of decisions: which features to prioritize in virtual screening, which machine learning architecture to use for prediction, which animal model to use for efficacy testing, how to design community engagement for clinical trials, which drug administration to prioritize given beliefs about patient adherence and healthcare infrastructure, and more. We can represent each team's overall scientific strategy as a vector of $N$ binary decisions, $d = (d_1, \ldots, d_N)$, where each element encodes a choice, such as whether to adopt a particular methodological assumption ($d_i = 1$) or not ($d_i = 0$).\footnote{Following works on $NK$ modeling, we treat these decisions as binary, without loss of generality.}

NK landscapes formalize how such decisions jointly determine the ``fitness'' of a research strategy, or how well its constituent decisions work together to achieve certain scientific goals. If fitness contributions of decisions were independent, optimization would be straightforward: evaluate each decision separately, identify the better option, and combine them. Scientific decision-making is rarely so simple, however, since the value of one choice often depends on others. For instance, adopting certain machine learning tools in healthcare settings may be beneficial, only when coupled with justified assumptions about the biological mechanisms and patient populations involved~\citep{chen2021ethical}.

In the $NK$ framework, the parameter $K$ formalizes this \textit{interdependence}. When $K = 0$, decisions do not interact, and contribute independently to overall fitness. Higher values of $K$ represent increasingly complex landscapes, where the contribution of one choice depends (on average) on $K$ other choices. In this case, for any focal decision $d_i$, its fitness contribution changes as a function of those other $K$ decisions.\footnote{In our model, the $K$ interdependencies of each decision $d_i$ are randomly generated for each landscape. Formally, the overall fitness of a given research strategy $d = \{d_1, \ldots, d_N\}$ is given by 
$f(d) = \frac{1}{N}\sum_{i=1}^N \phi_i\big(d_i; S(d_i)\big)$, where $S(d_i)$ represents the set of $K$ other decisions in $d$ whose values influence the contribution of $d_i$, $\phi_i$, and $\phi_i$ assigns a normalized payoff in $[0,1]$ to each combination of $d_i$ and decisions in $S(d_i)$. In our simulations, landscapes are regenerated for each run, and transformed and normalized so that $\max f(d) = 1$.}

The $NK$ model thus captures a central challenge of scientific problem-solving: researchers must navigate a landscape where altering a single theoretical or methodological commitment can help or hinder depending on their other choices. This explains why communities may become stuck on local optima: when $K$ is large, local improvements can mislead researchers about the location of global optima. 

\subsection{Epistemic Agents}
In practice, a research team's expertise and resources put constraints on which decisions it can independently evaluate and modify. In our drug discovery example, a computational biology team might specialize in virtual screening and molecular optimization, but lack expertise in clinical trial design or manufacturing scale-up. A health policy team might specialize in community engagement and health system integration, but require external guidance on computational methods. We model this variance in expertise, by assigning each agent, $a$, a \emph{specialization set}, $h_a$, that consists of a randomly generated subset of the total $N$ decisions, where $|h_a|=H < N$. While we keep the size of $H$ fixed for a given simulated community, the particular decisions that make up the specialization set differs between agents. The specialization set $h_a$ thus represents the decisions that agent $a$ can explore without external assistance. As $H$ decreases relative to $N$, agents face greater pressure to rely on external sources for navigating the full decision space. 

\subsection{Social Learning in Networks}\label{subsec:social}
One external source of guidance are scientific peers. Specifically, agents can learn from others to whom they are socially connected.\footnote{Here, we understand being a ``neighbor'' in terms of placement on a static, exogenously defined network structure.} We model scientific communities as random networks~\citep{erdHos1960evolution} where each pair of agents is connected with a fixed probability $p_{edge}$. To ensure that (cluster of) agents do not remain isolated, we sample these random networks under the constraint that they must be connected. Within this network, each agent engages in social learning with probability $p_{social}$. When doing so, it considers its neighbors and adopts the decision vector of the most successful one, provided that the solution improves its fitness.\footnote{In our reported results, we consider $p_{social} = \{0.1,0.2\}$. Sensitivity analyses varying this parameter reproduce the core findings of~\citet{lazer2007network}: higher $p_{social}$ accelerates short-term gains but reduces long-run exploration by pushing communities more quickly toward local optima.}

\subsection{The Computational Module and Problem Decomposability}\label{sec:modules}
Increasingly, research teams can also depend on AI tools as an external source of guidance. However, not all decisions facing a research team are equally amenable to AI assistance. In our drug discovery example, while decisions pertaining to data-analysis pipelines, chemical-property prediction, and simulation of biological processes may be amenable to computational tools, other decisions, such as those concerning aspects of animal studies, qualitative study design, or patient-adherence considerations may involve factors that are less amenable to automation.\footnote{Of course, AI systems might provide suggestions---potentially hallucinatory or unjustified ones---about a very many things. Here we are taking AI assistance to consist of something stronger akin to providing agents with the know-how to deliberate about certain decisions and enact change therein, even if those decisions are not part of their specialization sets.} To model this, we partition decisions into two modules: 
\begin{enumerate}
    \item \emph{Computational module} $m_c = \{d_1, \ldots, d_M\}$, consisting of decisions where AI tools can effectively operate.\footnote{We assume that the computational module includes the first $M$ of the $N$ decisions for ease of exposition and without loss of generality.} 
    \item \emph{Non-computational module} $\neg m_c = \{d_{M+1}, \ldots, d_N\}$ consisting of decisions outside AI's purview. 
\end{enumerate}

The relationship between decisions across the two modules raises interesting methodological considerations that are not simply about the number of dependencies (i.e., $K$), but the \textit{pattern} of those dependencies~\citep{ethiraj2004modularity,ganco2009nk,reijula2023division}: to what extent can the two modules be treated as separate or decomposable sub-problems? To make this point precise, for each decision $d_i \in d$, let $S(d_i)$ denote the set of $K$ other decisions whose values influence the fitness contribution of $d_i$. We can distinguish two sets of dependencies: 

\begin{itemize}
    \item $k_{\text{in}}(i)$: Number of \textit{within-module} dependencies, given by $|S(d_i) \cap m_c| +1$, if $d_i \in m_c$; and $|S(d_i) \cap \neg m_c| +1$, otherwise.\footnote{Notice that we add one to include the dependency of the fitness contribution of a decision on the decision itself. This allows $\rho = .5$ to be a meaningful middle point in problem modularity.}
    \item $k_{\text{out}}(i)$: Number of \textit{out-of-module} dependencies, given by$|S(d_i) \cap \neg m_c|$, if $d_i \in m_c$; and $|S(d_i) \cap m_c|$, otherwise. 
\end{itemize}

\noindent Within-module dependencies capture how a decision's value is influenced by all the decisions in the same module, whereas out-of-module dependencies capture how a decision's value is influenced by decisions in the other module. 

Tracking the proportion of within- and out-of-module dependencies for each module $m \in \{m_c, \neg m_c\}$ offers a way of tracking the extent of \textit{modularity} or \textit{decomoposability} of the two modules:
$$
\rho(m) := \frac{\sum_{d_i \in m} k_{\text{in}}(i)}{\sum_{d_i \in m} k_{\text{in}}(i) + \sum_{d_i \in m} k_{\text{out}}(i)} \in [0,1].
$$

\noindent A high $\rho(m)$ indicates that computational and non-computational modules form relatively self-contained subproblems. That is, in finding the optimal configuration of decisions in each module, one can largely abstract away from the configuration of decisions in the other. In contrast, a low $\rho(m)$ indicates that the contributions of decisions within the module depend heavily on decisions outside of it. In our running example, for instance, this can happen when the appropriateness of computational decisions depends critically on non-computational upstream (e.g., theoretical and methodological assumptions about data) or downstream considerations (e.g., qualitative patient study and community acceptance). Here, one might expect that optimizing only over the set of computational decisions without regards for the non-computational context can create solutions that are misaligned with the overall optimum. 

In our simulations, therefore, we vary $\rho(m)$ to explore how modularity moderates AI effectiveness, assuming symmetric modularity across both modules to isolate the effects of decomposability itself.\footnote{Of course, the dependency pattern need not be symmetric. For example, the dependencies in the non-computational module may form a relatively self-contained cluster, while the fitness contribution of decisions in the computational module may be impacted by out-of-module decisions (e.g., depending on technological and organizational factors influencing data collection). That said, the symmetry assumption enables us to focus on the extent (as opposed to potentially varied patterns) of decomoposability between modules.} In Sections~\ref{sec:simulation1-non-personalized} and~\ref{sec:simulation2-mitigation}, we consider different designs for AI systems, and describe implementation details about how they provide recommendations about the computational module. 

\subsection{Collective Epistemic Search}
We can now formalize collective problem-solving. Each agent $a$ is initialized with a random decision vector $d^a \in \{0,1\}^N$ and specialization set $h_a$. Agents know their current fitness, $f(d^a)$, and can evaluate alternatives. To obtain such alternatives, each round, agents pursue one of three strategies:

\begin{enumerate}
    \item \textbf{Social learning:} With probability $p_{social}$, the agent adopts the decision vector of their highest-performing neighbor, if it improves fitness.
    \item \textbf{Explore:} Otherwise, they can explore the landscape independently, by flipping a randomly selected decision from their specialization set $h_a$, adopting the change if it improves fitness.
    \item \textbf{Query AI:} For agents with access to AI tools, if they do not engage in social learning, instead of independent exploration, they can choose to query an AI tools, and receive recommendations for decisions within the computational module $m_c$, adopting them if they improve overall fitness. Agents' query behavior is determined by a community-level \textit{AI use rate} $\in [0,1]$. For example, in communities with an AI use rates of $0.25$, agents will, on average, seek AI recommendations in $25\%$ of rounds not spent on social learning. 
\end{enumerate}

In sum, each round: agents engage in social learning with probability $p_{social}$. When they aren't social learning, they query AI with a probability equal to the AI use rate, otherwise they explore. 
Importantly, each action consumes one round regardless of outcome. Inferior peer solutions, unsuccessful explorations, and rejected AI recommendations, all carry opportunity costs, reflecting that evaluating alternatives requires time and resources.

\section{Simulation 1: Non-personalized AI and The Formation of Epistemic Monocultures}\label{sec:simulation1-non-personalized}
Our first simulations examine the systemic impacts of adopting a single AI system that offers uniform best-practice recommendations. This ``non-personalized'' mode reflects typical deployments, where few pretrained systems provide potentially useful but context-insensitive guidance. When, and under what conditions, do such tools enhance collective problem-solving, and when do they instead produce detrimental epistemic monocultures?

\subsection{Experimental design}\label{subsec:}
\subsubsection{Formalizing Non-personalized AI (NP AI)}\label{subsec:non-personalized}
When agent $a$ queries the Non-personalized AI (NP AI) system, the tool returns the computational configuration used by the globally best-performing agent $b$. In this sense, NP AI acts akin to a social learning mechanism that is \emph{global} in reach but \emph{partial} in scope: it provides guidance about leading computational practices across all agents (not just the querying agent's local neighborhood), but it ignores how those practices interact with agents' non-computational choices.

This formalization captures some of the key dynamic underlying Messeri and Crockett's concern about monocultures of knowing (\S\ref{sec:background}): agents incorporate the ``best'' computational practices of successful peers without recognizing that those practices may be contingent on non-computational assumptions that do not necessarily hold in their own context.

\subsubsection{Dependent Measures}
We focus on two measures to evaluate the epistemic effects of adopting NP AI system.

\textbf{Epistemic success.}
We operationalize epistemic success as the average fitness of agents in the community, measured by the mean of $f(d^a)$ across all agents. Higher mean fitness indicates that the community, on average, has identified superior research strategies. We examine epistemic fitness at different points in a simulation run.

\textbf{Transient epistemic diversity.}
To capture diversity in research practices, we measure the \emph{mean pairwise Hamming distance} between agents' decision vectors at each round, and compute the area under this curve (AUC) over the course of a simulation. The Hamming distance between two binary decision vectors counts the number of positions on which they differ; a higher average distance thus indicates greater heterogeneity in the strategies pursued by agents. For example, two agents differing on three of ten decisions have a Hamming distance of 3. Tracking the mean Hamming distance across time provides a measure of \emph{transient diversity}---how varied the community's approaches are during its search for better solutions. Since our simulations converge when the community reaches a consensus, the mean Hamming distance in later rounds approaches zero. The AUC thus offers a useful summary statistic for how much diversity the community maintained before convergence. A larger AUC indicates that the community preserved its transient diversity for longer, while a smaller AUC reflects faster homogenization in research practices.\footnote{In all conditions reported, communities reach convergence well within the 100-round simulation window. By the end of the simulation, communities have converged to a single shared solution in nearly all replications (mean unique solutions  $\approx 1$ and mean pairwise Hamming distance $\approx 0$; for $100$ agents, over $1000$ simulation runs).}

We compare otherwise identical communities differing only in AI access across a range of parameter settings.\footnote{Specifically, we keep $N=20$ and $M=10$ fixed across all communities, while varying $K=\{5,9\}$, target modularity $\rho(m)$ from $0.6$ to $1$, $p_{social}=\{0.1, 0.2\}$, $H=\{5, 10, 15\}$, AI use rate $\in [0.125,0.875$] with $0.125$ increments. Note that since scientific problems are generally modular to some extent, we do not consider $\rho\leq.5$} Each configuration is simulated 1000 times.

\subsection{Results}
\begin{figure}[H]
    \centering
    \begin{subfigure}[b]{0.4\textwidth}
         \centering
         \includegraphics[width=\textwidth]{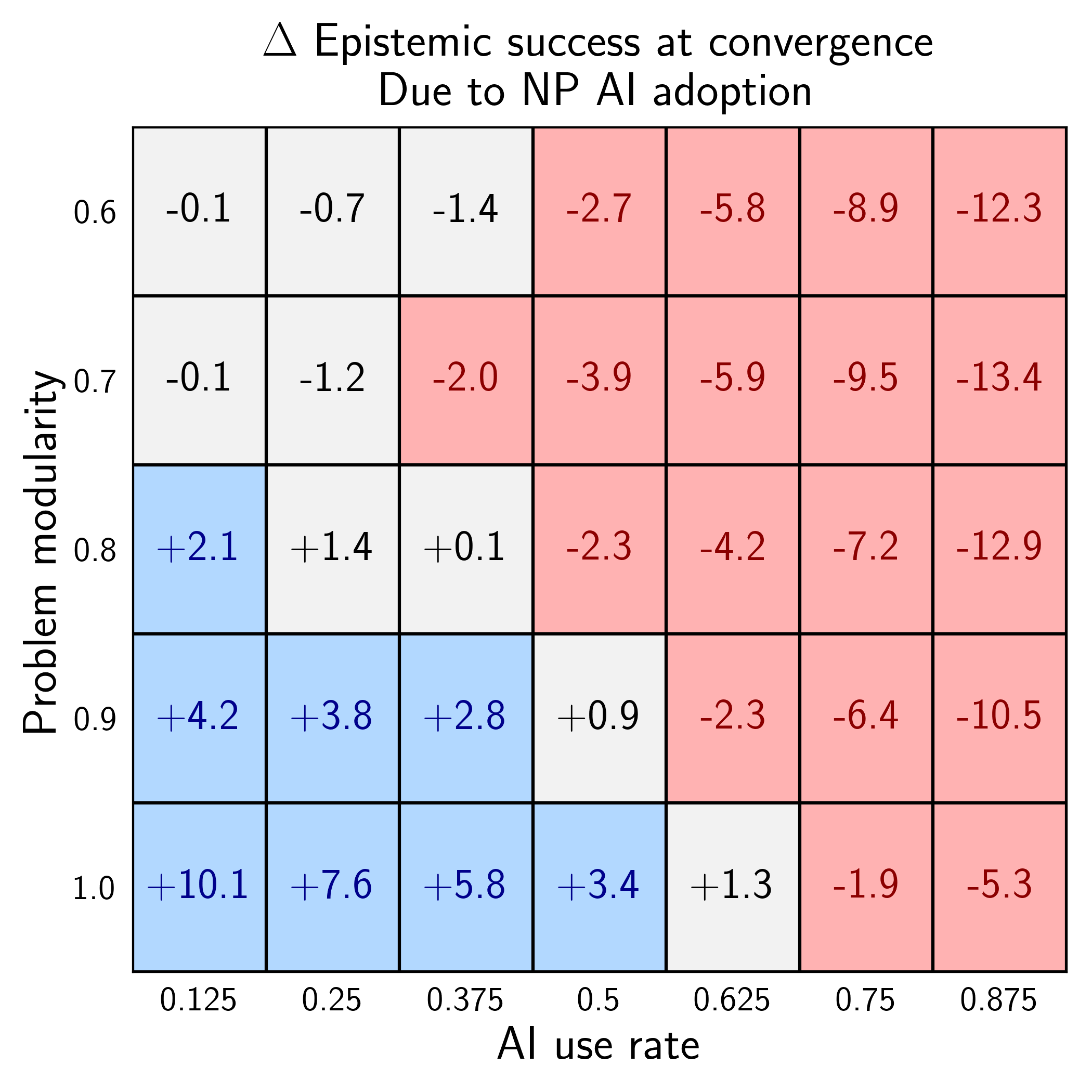}
         \caption{}
         \label{fig:non-personalized-fitness-H10-heatmap}
    \end{subfigure}
    \hfill
    \begin{subfigure}[b]{0.4\textwidth}
         \centering
         \includegraphics[width=\textwidth]{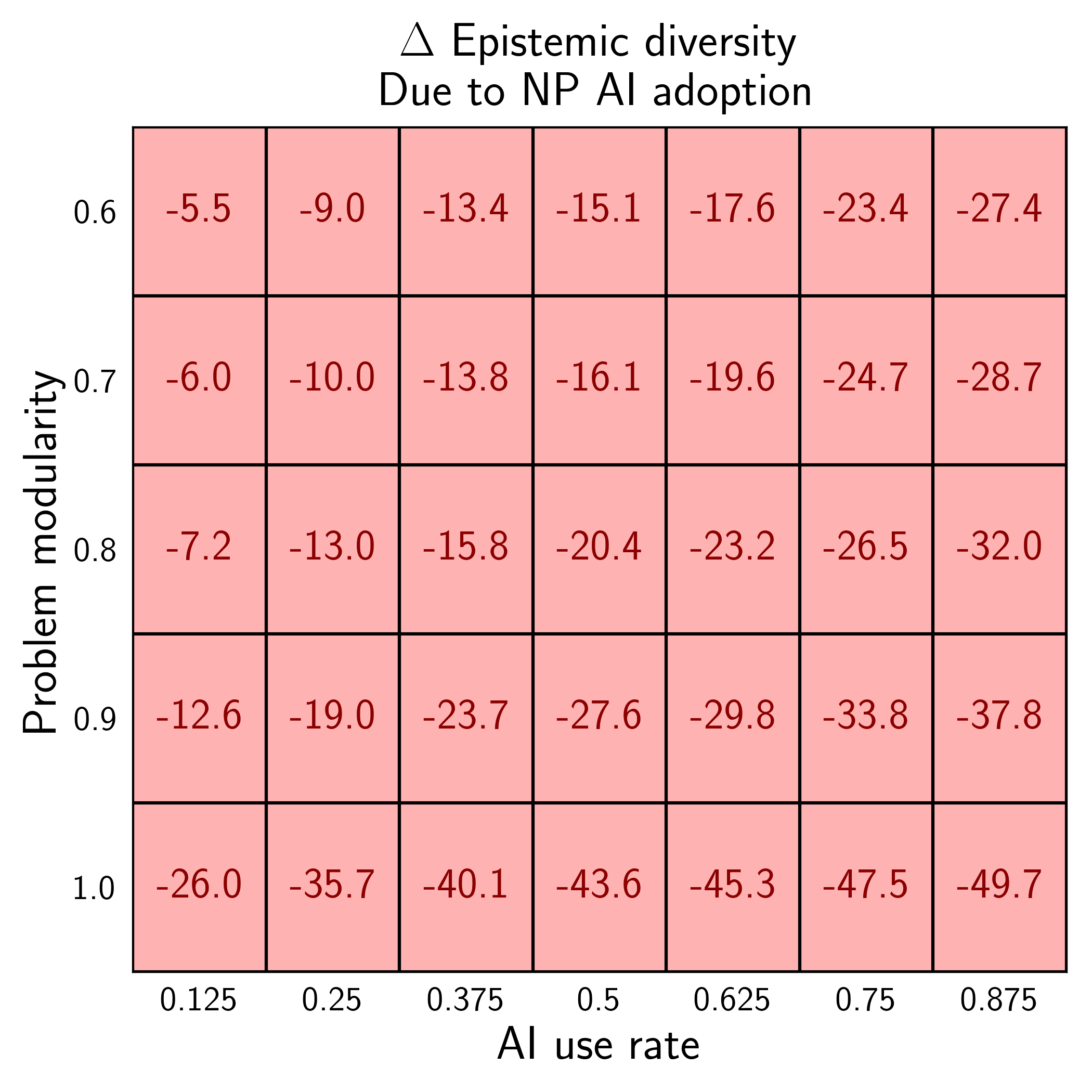}
         \caption{}
         \label{fig:non-personalized-diversity-H10-heatmap}
    \end{subfigure}
    \caption{The differential impacts of introducing Non-personalized (NP) AI, expressed as \textit{percentage differences} relative to communities without AI tools (averaged across 1000 simulation runs). Communities consist of $100$ agents with specialization set size $H=10$, navigating a problem with $N=20$, $K=9$, $M=10$, while engaging in social learning with $p_{social}=0.1$. Panel (A) shows differences in epistemic \textit{success} at convergence across problem modularity and AI use frequency. Panel (B) shows differences in transient epistemic diversity (measured by AUC of mean Hamming distance). Colored cells indicate statistically significant ($p < 0.05$) differences: \textcolor{blue}{blue} for improvements, \textcolor{red}{red} for declines, gray for non-significant ($p >= 0.05$).}
    \label{fig:non-personalized-AI}
\end{figure}

Our simulation results show that the impact of NP AI recommendations on epistemic success crucially depends on the extent of problem modularity. As shown in Figure~\ref{fig:non-personalized-fitness-H10-heatmap}, when the underlying problem is highly modular ($\rho(m)\ge0.8$), and can be decomposed into two relatively self-contained sub-problems, AI use---particularly at low to moderate levels---can yield reliable gains in community-level fitness relative to the no-AI baseline. In these settings, the problem structure allows AI's global ``best practice'' recommendations about $m_c$ configuration to efficiently inform agents' decisions within that module. 

These advantages disappear, however, as modularity decreases. In this case, the insensitivity of the AI tool to the heterogeneous contexts of agents---that is, their differing configurations in $\neg m_c$---adversely impacts its effectiveness. As Figure~\ref{fig:non-personalized-fitness-H10-heatmap} shows, when $\rho(m)\le0.7$, at convergence, NP AI provides negligible benefits at best, and becomes actively detrimental at high use rates. 

When it comes to transient epistemic diversity, we find that NP AI significantly reduces diversity across all settings by 25\% on average (See Figure~\ref{fig:non-personalized-diversity-H10-heatmap}). Importantly, however, this diversity loss is not \emph{necessarily} epistemically harmful. Rather, diversity reduction can reflect productive convergence when AI successfully coordinates the community toward superior solutions. This is illustrated by examining modularity's differential impacts on diversity and performance. At high modularity, NP AI reduces diversity substantially. Yet, this homogenization accompanies significant epistemic improvements, suggesting productive coordination around superior computational configurations. At low modularity, diversity loss is more modest, but occurs alongside performance stagnation or decline. 

\begin{figure}[H]
    \centering
    \begin{subfigure}[h]{0.35\textwidth}
         \centering
         \includegraphics[width=\textwidth]{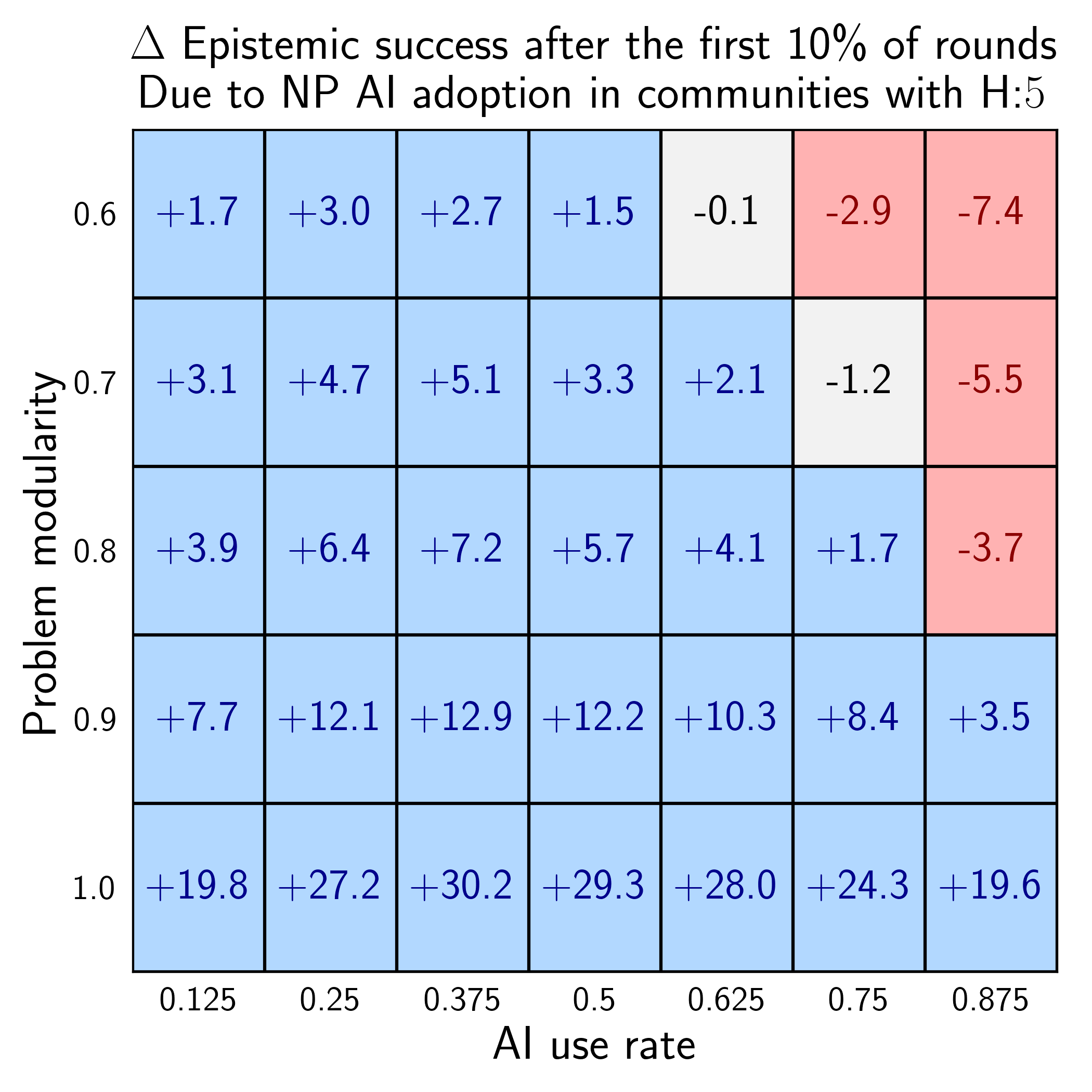}
         \caption{}
         \label{fig:non-personalized-early-fitness-h5}
    \end{subfigure}
    \begin{subfigure}[h]{0.35\textwidth}
         \centering
         \includegraphics[width=\textwidth]{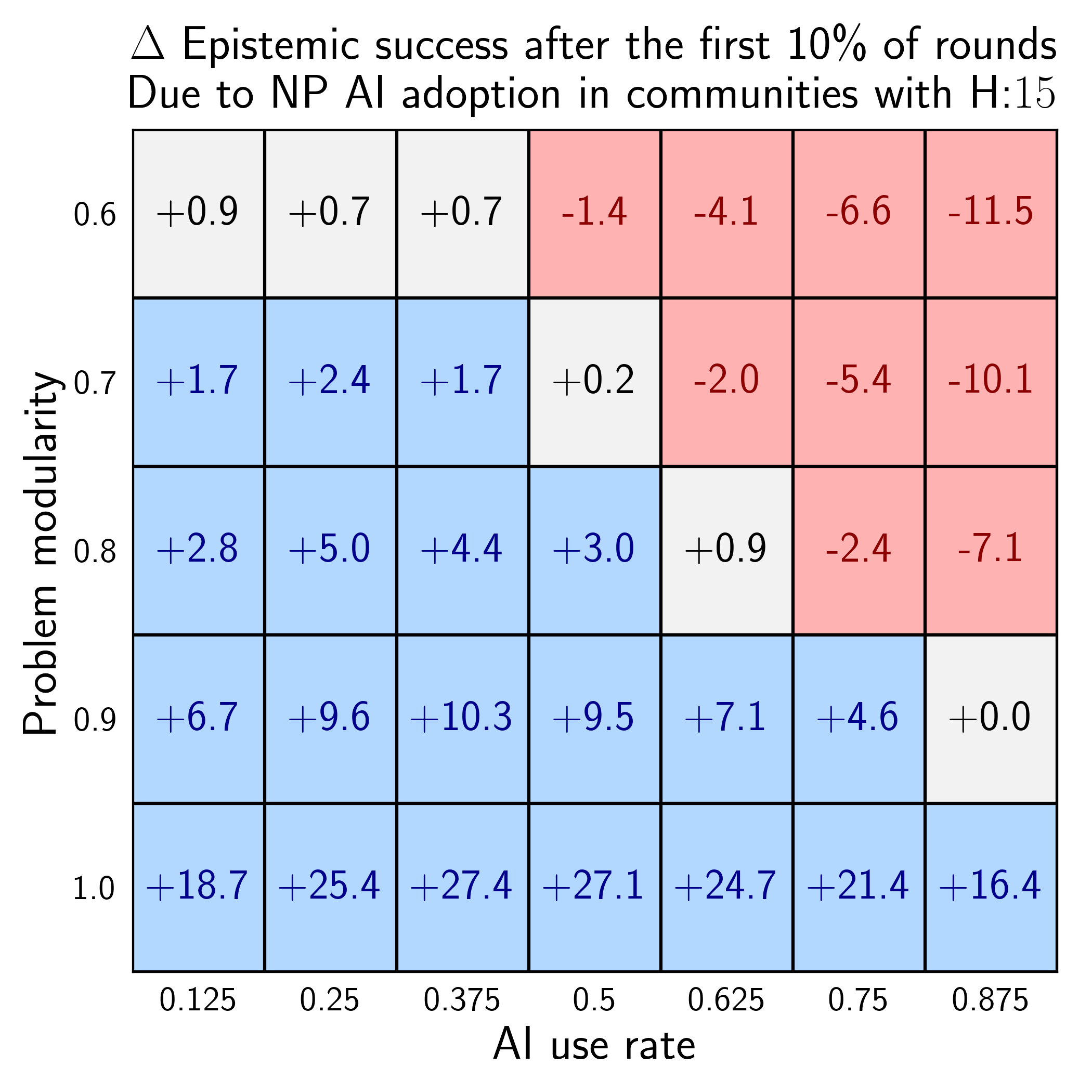}
         \caption{}
         \label{fig:non-personalized-early-fitness-h15}
    \end{subfigure}
    \hfill
    \begin{subfigure}[h]{0.35\textwidth}
         \centering
         \includegraphics[width=\textwidth]{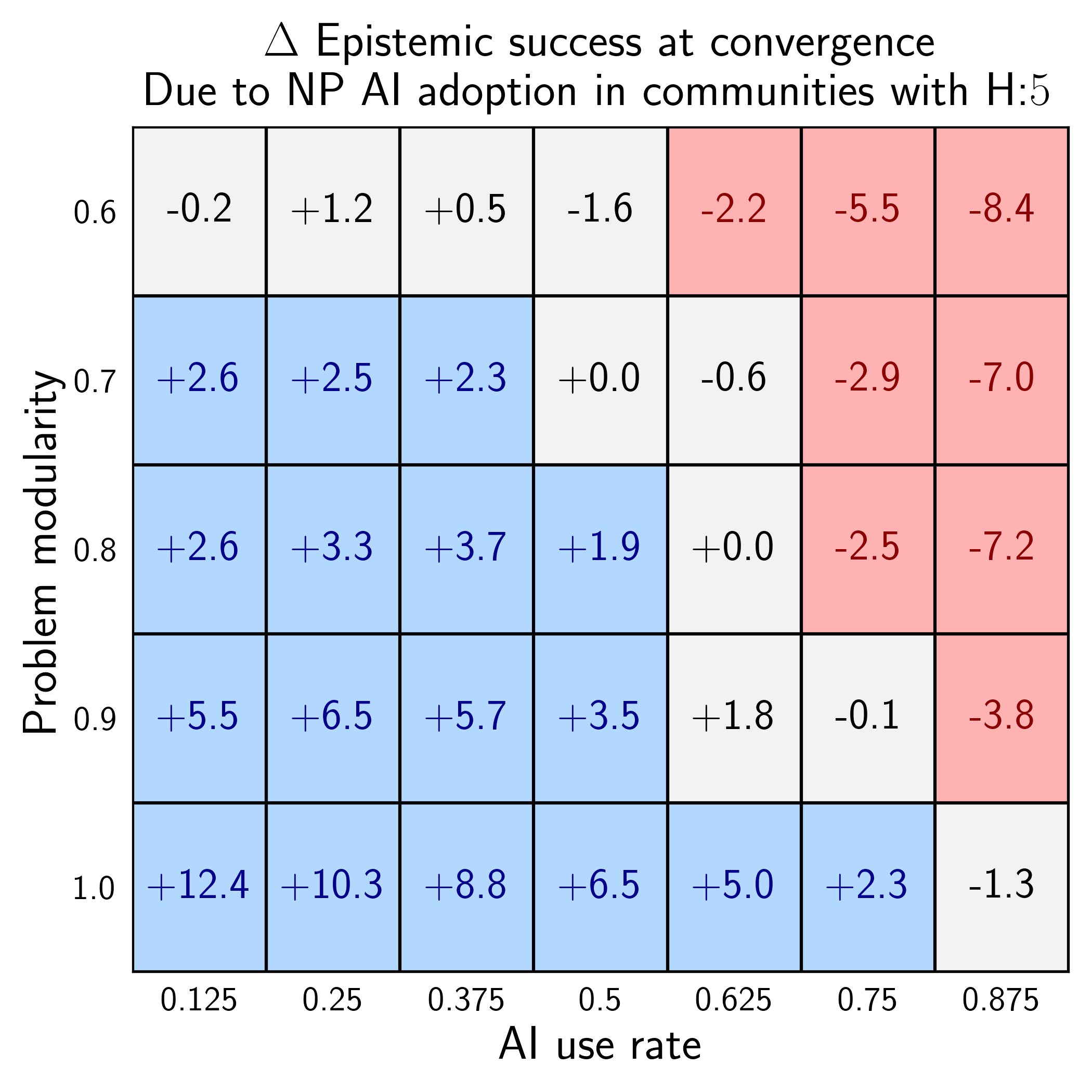}
         \caption{}
         \label{fig:non-personalized-final-fitness-h5}
    \end{subfigure}
    \begin{subfigure}[h]{0.35\textwidth}
         \centering
         \includegraphics[width=\textwidth]{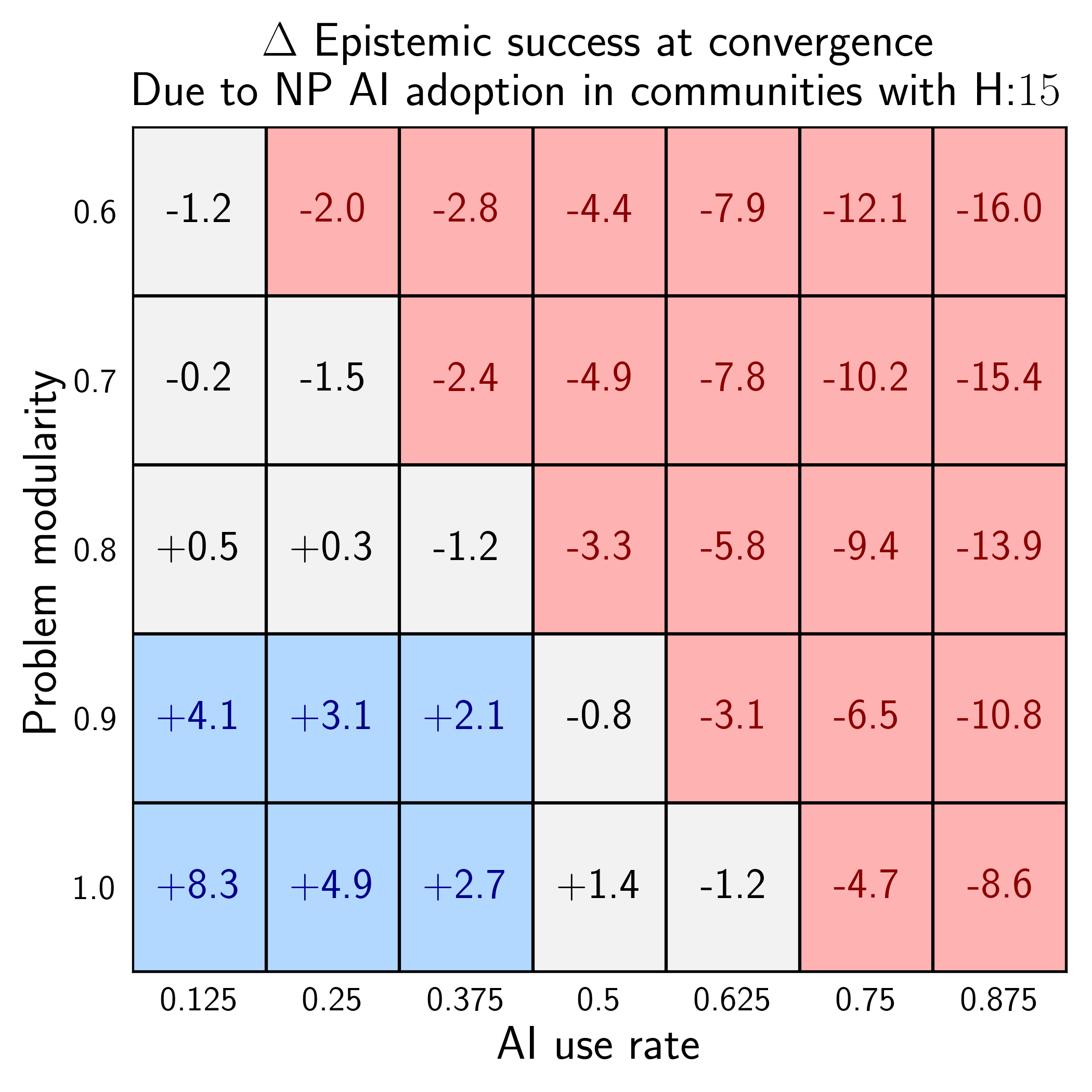}
         \caption{}
         \label{fig:non-personalized-final-fitness-h15}
    \end{subfigure}
    \caption{Difference in early versus final average epistemic success between communities with access to Non-personalized (NP) AI and communities without AI tools, expressed as percentage differences (averaged across 1000 simulation runs). Communities consist of $100$ agents facing a problem with $N=20$, $K=9$, $M=10$, while engaging in social learning with $p_{social}=0.1$. Panels (A) and (B) show percentage differences in \textit{early epistemic success} (first $10\%$ of rounds) across problem modularity and use rate for communities specialization set sizes $H=5$ and $H=15$, respectively; panels (C) and (D) shows percentage differences in final epistemic success after convergence for those communities. Colored cells indicate statistically significant ($p < 0.05$) differences: \textcolor{blue}{blue} for improvements, \textcolor{red}{red} for declines, gray for non-significant ($p \geq 0.05$).}
    \label{fig:non-personalized-AI-across-H}
\end{figure}

When, then, is homogenization harmful? Our analyses identify two mechanisms through which NP AI induces diversity loss, with differential epistemic consequences. First, especially in contexts with higher modularity or use rate, which increase uptake of AI recommendations, NP AI produces dynamics similar to those documented in studies of social learning~\citep{lazer2007network,zollman2010epistemic}. Specifically, NP AI recommendations based on the currently best-performing $m_c$ configuration provide early gains, but can cause communities to converge prematurely on local optima. This mechanism is reflected in the monotonic decrease of NP AI's benefits as use rate increases, even for highly modular problems (Figure~\ref{fig:non-personalized-fitness-H10-heatmap}).

Second, especially when lower modularity renders NP AI recommendations ineffective, the reliance on such systems can involve significant \textit{opportunity costs} due to context-mismatch of recommendations. In this case, the missed rounds of individual exploration, which were instead spent querying AI, lead to marked reduction in the diversity of practices in the community. This in turn limits the quality of solutions from which agents can socially learn. 

The marginal risk of both of these mechanisms of homogenization increases in more capable scientific communities that have more to gain from the potential benefits of transient diversity. When agents have small specialization sets ($H=5$), they explore limited regions of the decision space, leaving gaps that NP AI can fill. With larger specialization sets ($H=15$), agents collectively achieve broader coverage, and NP AI's marginal benefits may not justify its homogenizing costs. Figure~\ref{fig:non-personalized-AI-across-H} illustrates this pattern. NP AI offers substantial early gains regardless of specialization (Figures~\ref{fig:non-personalized-early-fitness-h5} and ~\ref{fig:non-personalized-early-fitness-h15}). After convergence, however, while communities with $H=5$ retain small but significant benefits (Figure~\ref{fig:non-personalized-final-fitness-h5}), those with $H=15$ frequently see early gains reverse into losses (Figure~\ref{fig:non-personalized-final-fitness-h15}). Accordingly, while coordinating around NP AI recommendations accelerates initial progress, it leads to premature convergence---a speed-performance tradeoff mediated by diversity loss, familiar from prior work in philosophy of science~\citep{zollman2010epistemic, lazer2007network}.

Overall, our findings reveal that NP AI's effectiveness depends on a specific alignment of problem structure (high modularity), practices (moderate use), and existing research capabilities (limited specialization creating coverage gaps). Outside these conditions, NP AI harms collective performance by reducing valuable transient diversity without providing compensating coordination benefits, supporting Messeri and Crockett's concerns about AI-instigated epistemic monocultures.\footnote{We observe qualitatively similar patterns concerning the results discussed in the section for $K=5$, and $p_{social}=0.2$ as well.}
\section{Simulation 2: Mitigating Epistemic Monocultures with Randomization and Personalization}\label{sec:simulation2-mitigation}
Our second set of simulations explores two design interventions for mitigating the homogenization documented in Section~\ref{sec:simulation1-non-personalized}: \emph{randomizing} recommendations and \emph{personalizing} them. Randomization introduces diversity by sampling from multiple high-performing options rather than always recommending the single best, while personalization tailors recommendations to each agent's context. Below, we detail the implementations, and describe our findings about the conditions under which each approach succeeds. As before, we keep the broader discussion for Section~\ref{sec:discussion}.

\subsection{Experimental Design}
\subsubsection{Formalizing Randomized Non-personalized AI (TDR AI)}
We implement a design inspired by randomization as a strategy for combating outcome homogenization~\citep{jain2024position}. The idea is that instead of recommending the same ``best'' option to everyone, the system randomizes over high-performing options, potentially preserving heterogeneity, while maintaining utility. We operationalize this as follows: when an agent queries the AI tool, the tool randomly selects an agent from the top decile performers in the community and returns the computational portion of its decision vector. We will refer to this implementation as Top-Decile Randomized AI, or TDR AI for short. This approach is non-personalized, insofar as it only considers the state of the community at the time of the query, ignoring the querying agent's context. In contrast with non-personalized AI design above, however, it samples from high performers (as opposed to only the single best performer), thus promoting diversity, while ensuring a high baseline of quality. 

\subsubsection{Formalizing Personalized AI (SBP AI)}
Personalized AI has been proposed as a way to mitigate the homogenizing effects of AI adoption~\citep{fugener2021will}. We operationalize this idea as follows. The AI tool identifies which \emph{single bit} in the computational subspace, $m_c$, the agent $i$ should flip in its current research practice $d^i$ to yield the greatest improvement. We will refer to this implementation as Single-Bit Personalized AI, or SBP AI for short.\footnote{We also consider an alternative personalized design which evaluate each community member's computational module in the context of the querying agent's non-computational decisions, and recommends the one that yields the highest fitness for the querying agent. See the online-only appendix for details.}

This design can be seen as a \emph{personalized myopically greedy} procedure: It models systems that guide researchers toward the next most promising experiment or methodological change \textit{for them}. This is unlike the non-personalized designs above, which simply recommend the computational portion of best or top-performing agents' practices without taking into account AI querying user's existing non-computational decisions. This approach thus models a highly tailored, context-sensitive use of AI that optimizes its recommendations about where to intervene next for each specific user's context. This is a demanding and highly idealized design, and we return to the assumptions that are implicit in this conception in Section~\ref{sec:discussion}.\footnote{We also conducted sensitivity test with an error prone version of SBP AI, which suggests the best bit to change with a certain level of accuracy ($1-\epsilon$), suggesting a random bit otherwise ($\epsilon$). The system performs slightly worse, but did not change the qualitative patterns discussed below.}

If the two mechanisms of homogenization identified in Section~\ref{sec:simulation1-non-personalized} are correct, then we should have differential expectations about the effectiveness of these mitigation strategies. The randomization implemented in TDR AI should address premature convergence when NP AI is effective (at high modularity) by diversifying which recommendations agents receive, and preventing universal adoption of a single configuration. However, it cannot resolve context-mismatch problems and the opportunity cost of using NP AI in low modularity settings. In contrast, the personalization afforded by SBP AI should address both mechanisms: by tailoring recommendations to individual contexts, it increases recommendation  effectiveness (addressing opportunity costs), and by providing different recommendations to different agents, it naturally generates diverse search trajectories (preventing premature convergence). We examine the systemic effects of these designs through the same experimental parameters as Section~\ref{sec:simulation1-non-personalized}.

\begin{figure}[H]
    \centering
    \begin{subfigure}[b]{0.35\textwidth}
         \centering
         \includegraphics[width=\textwidth]{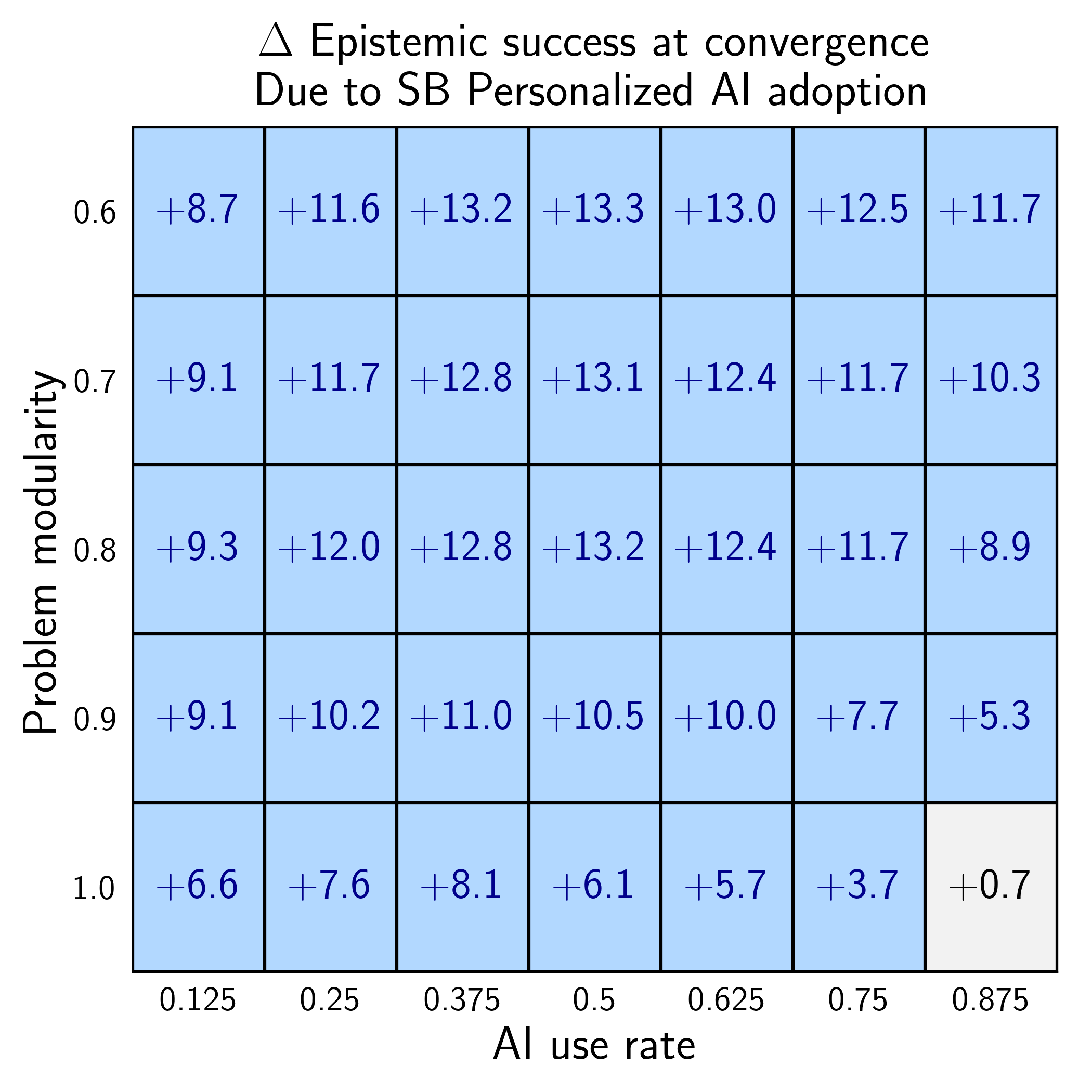}
         \caption{}
         \label{fig:bc-fitness-heatmap}
    \end{subfigure}
    \begin{subfigure}[b]{0.35\textwidth}
         \centering
         \includegraphics[width=\textwidth]{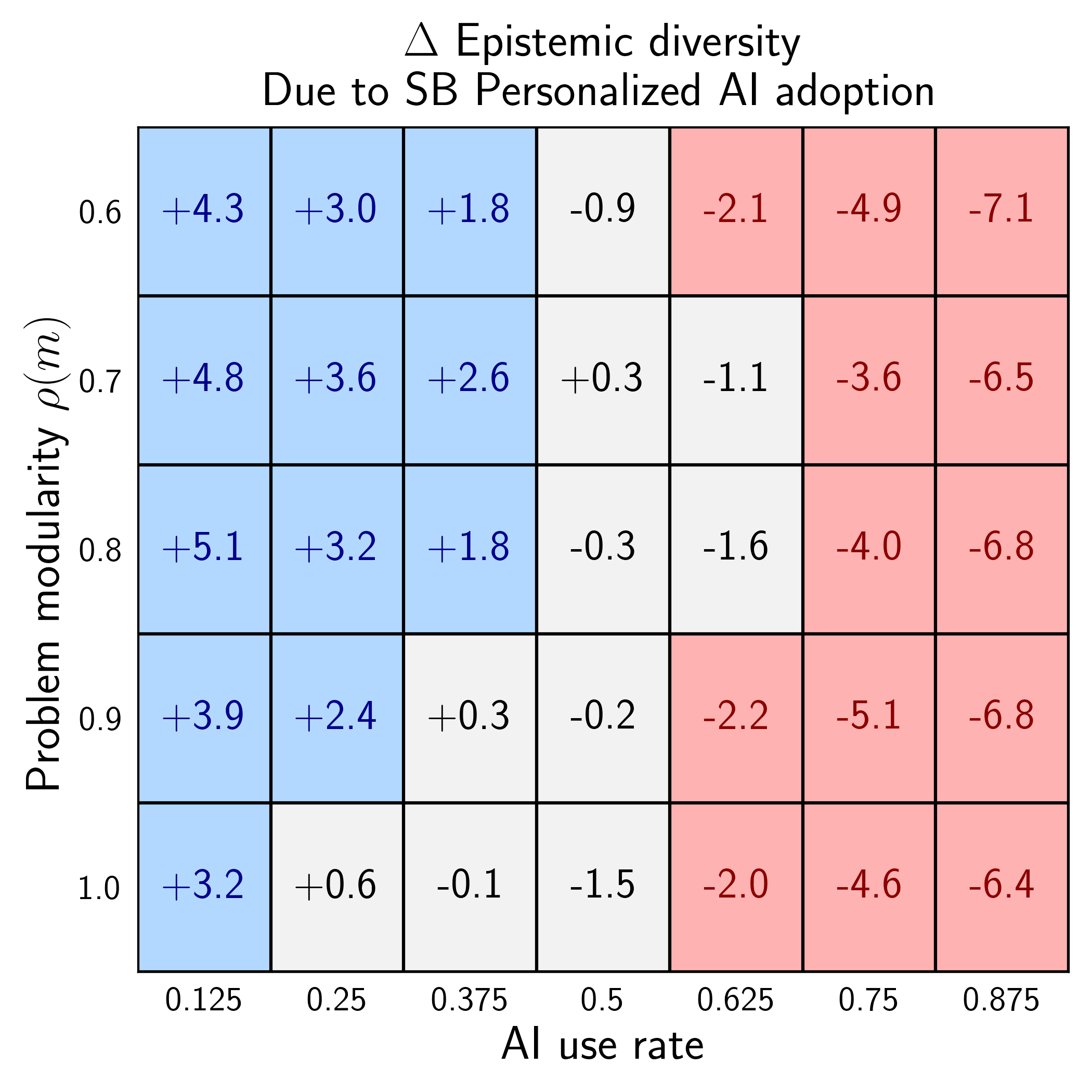}
         \caption{}
         \label{fig:bc-diversity-heatmap}
    \end{subfigure}
    \hfill
    \begin{subfigure}[b]{0.35\textwidth}
         \centering
         \includegraphics[width=\textwidth]{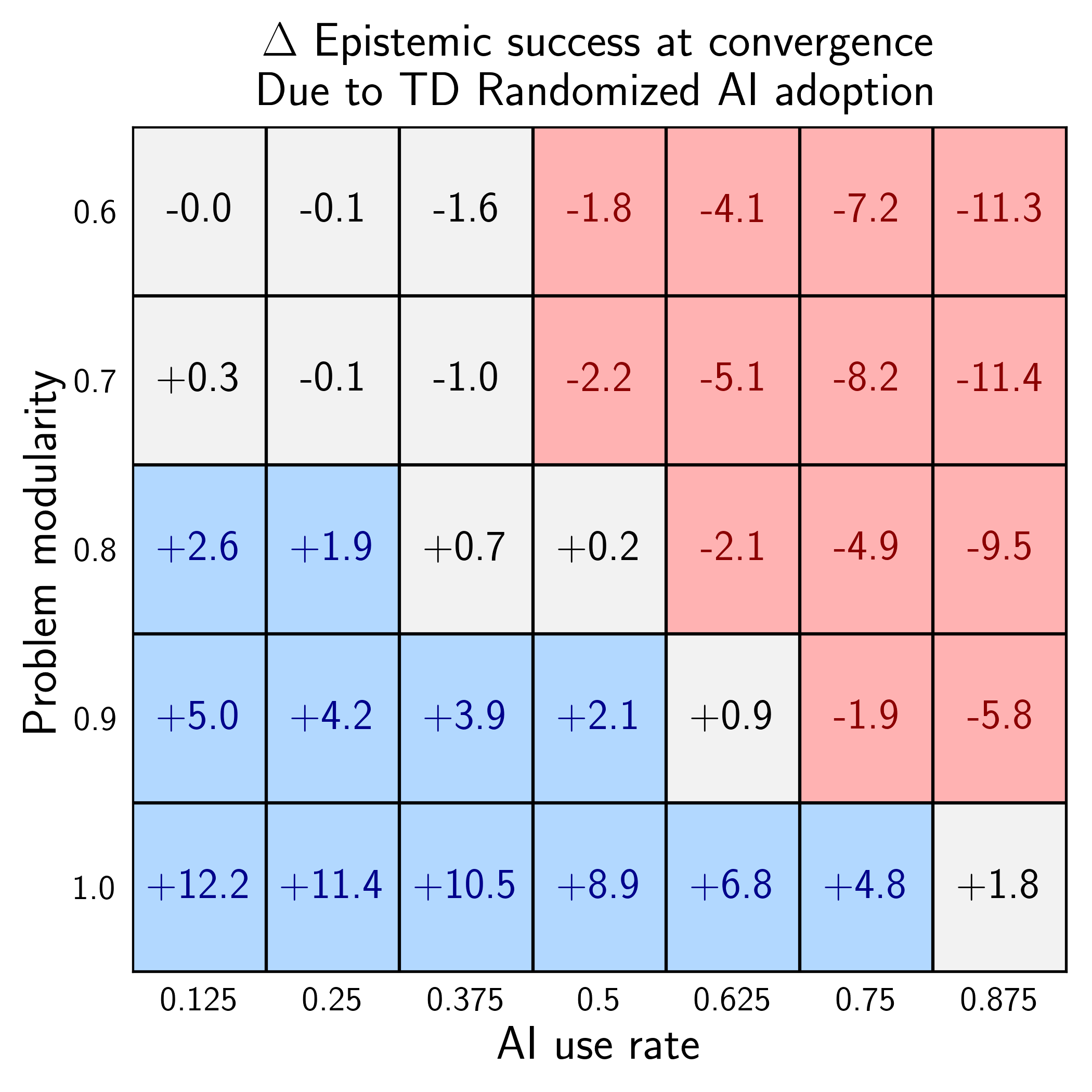}
         \caption{}
         \label{fig:td-fitness-heatmap}
    \end{subfigure}
    \begin{subfigure}[b]{0.35\textwidth}
         \centering
         \includegraphics[width=\textwidth]{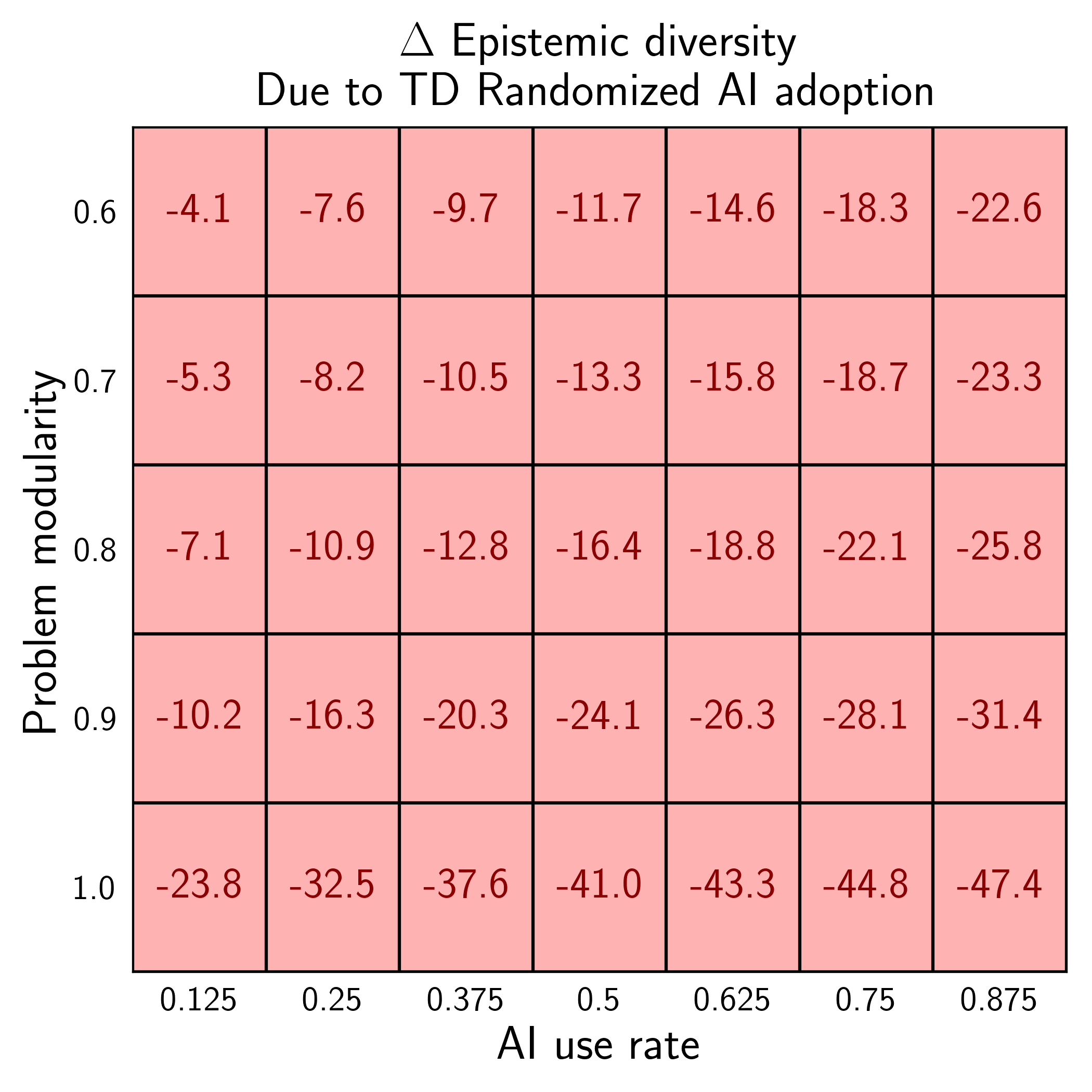}
         \caption{}
         \label{fig:td-diversity-heatmap}
    \end{subfigure}
    \caption{Percentage differences in average outcomes between communities with access to AI tools and communities without AI (averaged across 1000 simulation runs). Communities consist of $100$ agents with $H=10$, facing a problem with $N=20$, $K=9$, $M=10$, and $p_{social}=0.1$. Panel (A) and (B) show \textit{Single-Bit (SB) Personalized AI} effects on epistemic success at convergence and transient diversity, respectively, across problem modularity and AI use rates; panel (C) and (D) show these effects for \textit{Top-Decile (TD) Randomized AI} across the same conditions. Colored cells indicate statistically significant ($p < 0.05$) differences: blue for improvements, red for declines, gray for non-significant ($p \geq 0.05$).}
    \label{fig:mitigation-AI}
\end{figure}

\subsection{Results}
Our simulations reveal that personalization can enable benefits across structural conditions where non-personalized approaches fail. Figure~\ref{fig:bc-fitness-heatmap} demonstrates SBP AI's most striking advantage: in conditions where Non-personalized AI proves ineffective or harmful, SBP AI provides substantial improvements. This pattern validates our theoretical prediction that, by tailoring computational recommendations to each agent's specific non-computational configuration, SBP AI avoids the cross-module interference that undermines the effectiveness of non-personalized recommendations in tightly-coupled problems.

SBP AI's context-sensitivity has implications for epistemic diversity as well. We find that, especially at lower use rates, SBP AI maintains or even slightly increases transient diversity, compared to no-AI baselines (see Figure~\ref{fig:bc-diversity-heatmap}). This occurs because tailored recommendations guide agents toward \textit{different} (compared to non personalized AI) and \textit{more varied} (compared to non personalized AI as well as baseline communities dependent only on their own specialization sets) promising solutions. This diversity-preserving property stands in stark contrast to our non-personalized 
approaches, which reduce diversity across all conditions (see Figures~\ref{fig:non-personalized-diversity-H10-heatmap} and~\ref{fig:td-diversity-heatmap}).

As shown in Figure~\ref{fig:td-diversity-heatmap}, TDR AI offers no such benefits at low modularity. In this context, randomization performs comparably to SBP AI, with both approaches ranging from ineffective to actively harmful as use rates increase. This similarity confirms that the fundamental limitation of non-personalized recommendations is not just due to a lack of variety in recommendations, but additionally their insensitivity to individual contexts. Randomizing among top performers merely varies \emph{which} context-mismatched recommendation each agent receives without addressing the underlying structural challenge.

Interestingly, however, the comparative effectiveness of these mitigation strategies somewhat changes, under high modularity. As Figures~\ref{fig:bc-fitness-heatmap} and~\ref{fig:td-fitness-heatmap} show, when $\rho(m)=1$, TDR AI not only maintains the benefits of non-personalized AI at higher query rates; it even slightly outperforms SBP AI across all use rates. Notably, this advantage emerges despite randomization's non-personalized nature. At high modularity, module independence renders context-sensitivity less critical, and diverse computational configurations can succeed paired with various non-computational choices. Under these conditions, the diversity introduced by randomization sustains the benefits of simple non-personalized AI, even at higher use rates.

What accounts for the relative reduction in SBP AI's benefits in fully decomposable problems with $\rho(m)=1$? We can think of two related mechanisms. First, at $\rho(m)=1$, each module operates as a highly complex but independent subproblem with multiple high-quality local optima. This can be because SBP AI recommendations myopically guide agents along greedy paths toward whichever local optimum is nearest to their current position within this module. Given the higher internal complexity of the computational module in modular problems, this can more easily result in agents getting stuck in different local optima. Indeed, we find that increased modularity has a detrimental impact on the performance of baseline (no AI) community. But, as Figure~\ref{fig:bc-fitness-heatmap} shows, the drop in performance is even more substantial for the SBP AI. This may be because in the no AI community, the alternative computational decisions that the agents adopt through exploration are simply \textit{better} than their current ones, preserving some transient diversity, whereas with SBP AI agents adopt the myopically \textit{best} decision change for the entire computational module~\citep[For a similar phenomenon see][]{wu2024better}. 

Second, problem decomposability can fundamentally alter the efficiency of division of labor in collectives relying on SBP AI. At lower modularity, tight coupling between modules means that finding good configurations in $m_c$ also helps in exploring the fitness contributions of decisions in $\neg m_c$. In perfectly decomposable problems, however, these benefits disappear. Module independence means agents with identical $m_c$ configurations always receive identical recommendations, regardless of differences in $\neg m_c$. SBP AI recommendations thus reveal little about the non-computational module, and time agents allocate to querying AI---or to exploring computational decisions on their own---represents opportunity cost for optimizing the independently-complex non-computational module. This results in a defective division of cognitive labor across the two subproblems. 

These two mechanisms suggest different enhancement strategies. If the issue is due to the myopic implementation of SBP AI, then it can be addressed by a less myopic version of personalization that better deals with complexity, at least in contexts when this is technologically viable. If the issue is due to a defective division of cognitive labor, then the solution is not merely technological, but requires a restructuring of existing practices. 

\begin{figure}[t]
    \centering
    \begin{subfigure}[b]{0.32\textwidth}
         \centering
         \includegraphics[width=\textwidth]{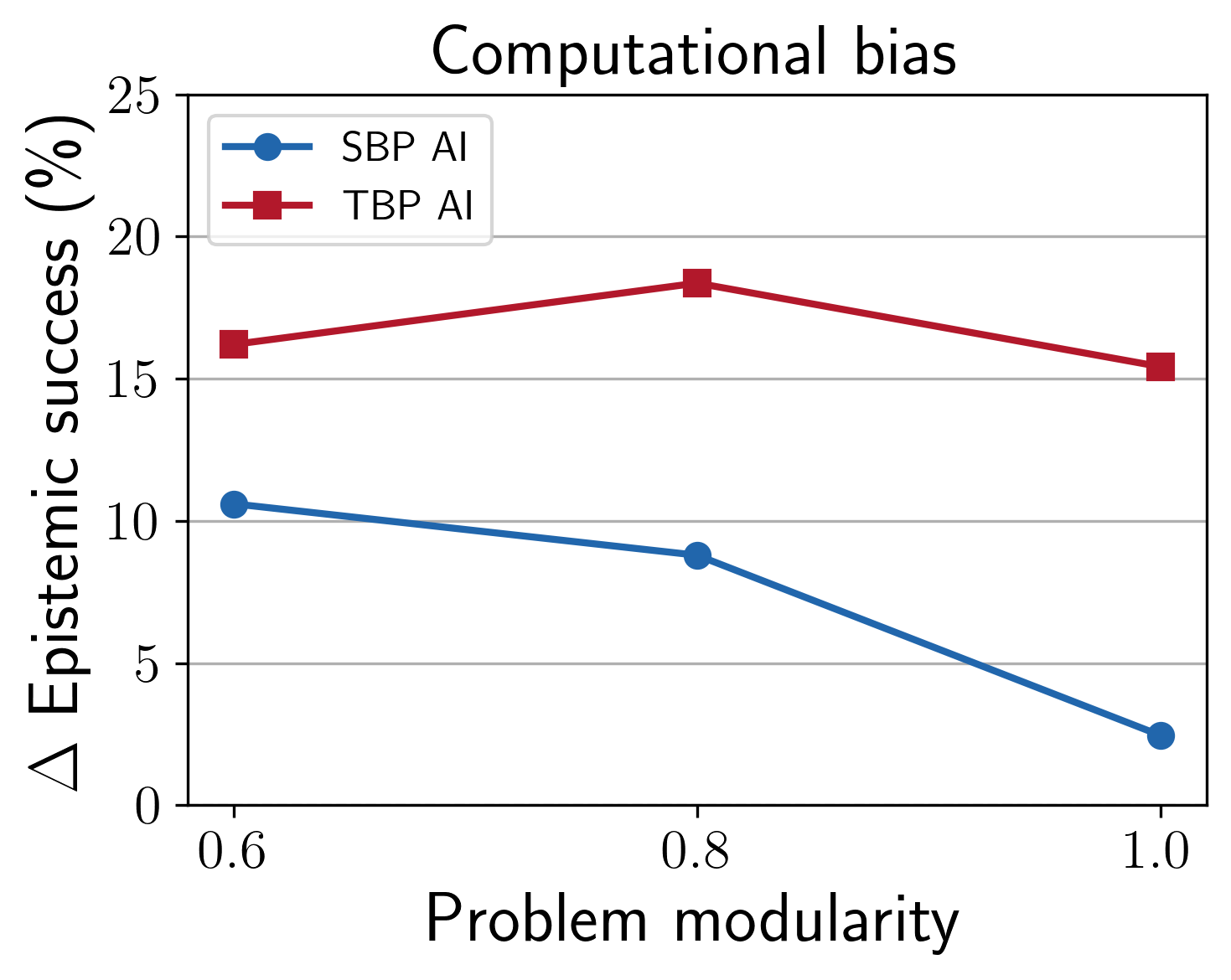}
         \caption{}
         \label{fig:computational}
    \end{subfigure}
    \hfill
    \begin{subfigure}[b]{0.32\textwidth}
         \centering
         \includegraphics[width=\textwidth]{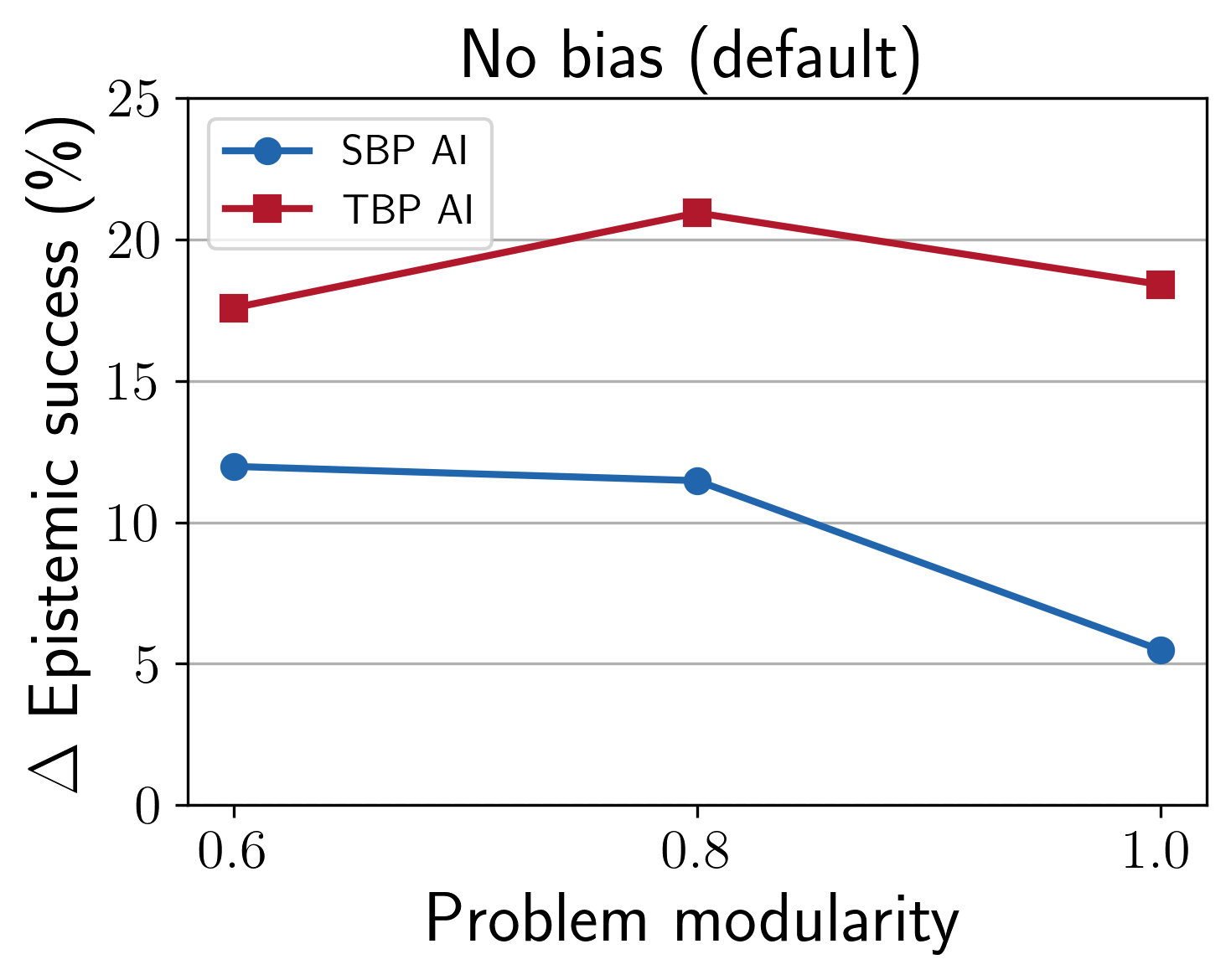}
         \caption{}
         \label{fig:no-bias}
    \end{subfigure}
    \hfill
    \begin{subfigure}[b]{0.32\textwidth}
         \centering
         \includegraphics[width=\textwidth]{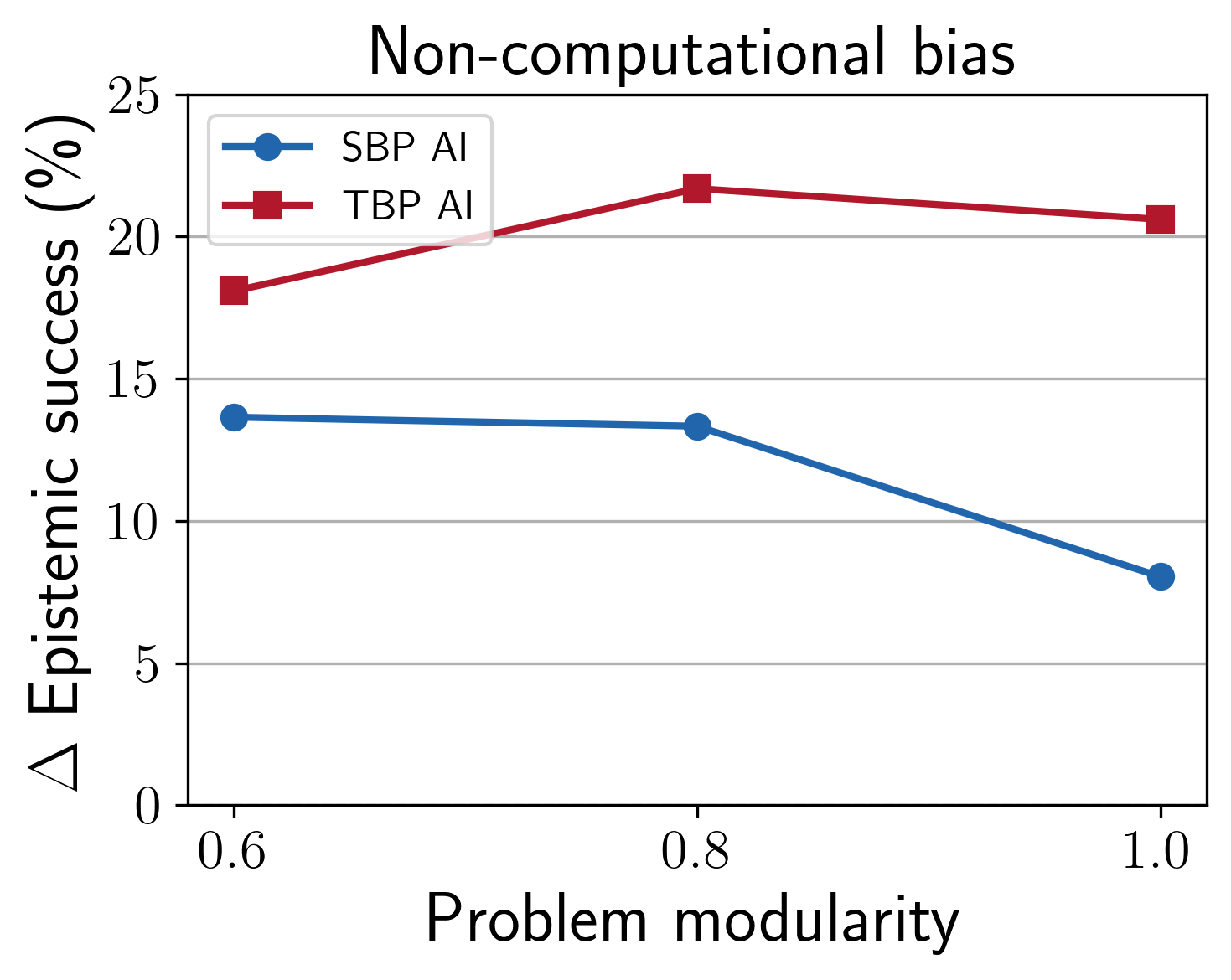}
         \caption{}
         \label{fig:non-computational}
    \end{subfigure}
    \caption{Percentage differences in epistemic success at convergence between communities with access to two variants of personalized AI tools, compared to communities without AI (averaged across 1000 simulation runs and across AI use rates): \textcolor{blue}{blue} lines show the difference due to adopting \textit{Single-Bit Personalized (SBP) AI}, while \textcolor{red}{red} lines show the difference due to adopting \textit{Two-Bit Personalized (TBP) AI}. Communities consist of $100$ agents with specialization set size $H=10$, navigating a problem with $N=20$, $K=9$, $M=10$, while engaging in social learning with $p_{social}=0.1$. Each panel shows one exploration bias condition: agents' post-AI-adoption search practice is biased toward (A) the \textit{computational} module, (B) unbiased (our default thus far), or (C) biased toward the \textit{non-computational} module.} 
    \label{fig:personalized-AI-exploration-bias}
\end{figure}

To further explore this idea, we examined two further variants of the model above. First, we consider a two-bit version of the Personalized AI system (TBP AI), which recommends to agents the optimal \textit{two bits} to change in the computational subspace (as opposed to the single bit case above). Intuitively, this means the system has a higher capability of guiding agents through complex problems. Second, we also explored what happens when communities \textit{alter} their exploratory practices following AI adoption. Specifically, we compared communities where agents bias their individual exploration toward computational decisions (reinforcing AI's focus), toward non-computational decisions (complementing AI's scope), or explore uniformly without bias (our default condition in prior simulations).\footnote{We implement exploration bias by adjusting how agents' specialization sets $h_a$ are drawn: computational bias places $\lfloor 0.75\,H\rfloor$ of the $H$ bits in $m_c$, subject to module-size constraints, and the remainder in $\neg m_c$ (non-computational bias does the reverse), while unbiased exploration---our default condition thus far---draws uniformly from all $N$ decisions.}

The results, shown in Figure~\ref{fig:personalized-AI-exploration-bias}, provide support for the mechanisms suggested above. Unsurprisingly, the more capable TBP AI not only outperforms the SBP AI, across different exploration biases; it can also perform \textit{better} with increased modularity (compare $\rho(m)=0.6$ with $\rho(m)=1$ in Figures~\ref{fig:no-bias}~\ref{fig:non-computational}). This makes sense given that increased modularity negatively impacts the performance of the baseline, no-AI community. What is more, for both SBP and TBP implementations, exploration bias moderates their effectiveness. When agents bias exploration toward non-computational decisions (Figure~\ref{fig:non-computational}), performance improvements increase markedly across all modularity levels compared to computational bias (Figure~\ref{fig:computational}), and no bias (Figure~\ref{fig:no-bias}) with more pronounced effects in fully decomposable problems. 

These results demonstrate that, apart from existing technological capabilities that can inform the effectiveness of AI systems in dealing with longer term problem complexity, Personalized AI's effectiveness can also depend on how communities adapt their norms and practices to AI adoption. Productive AI-human division of labor emerges not from passive AI adoption, but from organizational adjustments that direct human effort toward complementary problems AI cannot address. We now turn to these broader implications.

\section{Discussion}\label{sec:discussion}
Our simulations revealed four high-level results: First, Non-personalized AI's epistemic value is fundamentally conditional. Uniform recommendations yield benefits only under a narrow conjunction of highly modular problem structure, moderate use rates, and limited baseline capabilities. Outside these boundaries, Non-personalized AI ranges from ineffective to actively harmful. Second, Non-personalized AI systems can trade coordination gains against transient diversity, with varied epistemic consequences. At high modularity, substantial diversity loss accompanies performance improvements, reflecting productive convergence. At low modularity, diversity loss occurs without compensating longer term gains, indicating premature convergence on and opportunity costs due to context-inappropriate solutions. Third, Randomization---our implementation of a proposed technical mitigation---fails to address the underlying issue of context-mismatch. We find its utility is restricted to the same narrow structural conditions that enable Non-personalized AI, offering no benefit for complex, tightly coupled problems. Fourth, Personalization---our implementation of another proposed mitigation---enables robust benefits across structural conditions, while often preserving or enhancing diversity. Yet, realizing its full benefits may depend on organizational adaptation and coordinated adjustments in shared epistemic practices.

Overall, the model does not only help assess claims about AI-driven homogenization; It also allows us to explore the effectiveness of mitigation strategies proposed in the AI literature, namely randomization among high-performing recommendations and personalization. These interventions are most naturally interpreted as design features of AI systems, rather than of other drivers of homogenization. Our findings about them are therefore especially relevant to the sociotechnical design and governance of AI tools in science.

\subsection{From Tool Adoption to Institutional Preparedness and Adaptation}
Perhaps the key theme emerging from our experiments is that AI's impact on collective inquiry depends critically on how epistemic communities \textit{adapt} their norms and practices to exploit AI's capabilities, while compensating for its limitations. This distinction between passive adoption and active adaptation emerges most clearly when considering the structural prerequisites for different AI designs and the institutional interventions that moderate their risks and benefits.

\subsubsection*{Non-personalized AI: Modularity, Randomization, and Use Moderation}
Consider first the factors that determine and moderate the impacts of Non-personalized AI. Our findings indicate that the effectiveness of the non-personalized AI systems of the sort modeled here may be restricted to highly modular problems with moderate use rates. Outside these conditions, context-insensitive recommendations may provide little utility or, worse, accelerate premature consensus around inferior solutions. This restriction carries important organizational implications.

It underscores the need for careful structural assessment before deploying such AI systems. In particular, our results suggest that institutions may want to evaluate whether target problems are sufficiently decomposable for uniform AI guidance to succeed. Such assessments would require determining abstraction boundaries---identifying which factors fall within the purview of a problem, and which aspects can be safely treated as independent. This determination is complex and often contested. Recent debates about AI integration in high-stakes decision-making illustrate these challenges~\citep{selbst2019fairness,fazelpour2020algorithmic}. These debates reflect deeper disagreements about whether socially desirable objectives in AI-based decision-making can be framed computationally, particularly in ways that are meaningfully decomposable from human, organizational, and social factors, and thus illustrate the challenge of decomposing complex social problems into clearly separable computational and non-computational modules. 

Even in sufficiently decomposable problems, the benefits of non-personalized AI guidance crucially depends on appropriately navigating the speed-performance (or efficiency-diversity) trade-offs. Our results suggest both technical and institutional interventions. The success of TD Randomized AI systems in modular settings, for example, provides supports for both technical interventions, such as injecting structured diversity, and social interventions, such as fostering a pluralistic model ecosystem.\footnote{Of course, these approaches can differ along other dimensions of benefits and challenges. So, the overall choice, if one needs to be made, depends on those other considerations.} 
Moreover, our results suggest an important role for institutionalizing appropriate usage norms (e.g., via query budgets) to limit excessive reliance on non-personalized systems and avoid premature lock-in.

\subsubsection*{Personalized AI: Context Legibility and Complementarity}
Personalized AI systems introduce distinct requirements. Even beyond the technical prerequisites which may not be realistic in many domains, a necessary \textit{institutional} precondition for personalized AI is \textit{context legibility}. In our model, the SBP and TBP AI systems have perfect access to an agent's decision vector. In practice, however, making a research team's context legible to AI systems (or even to themselves and other teams) requires massive institutional work, including the creation of explicit protocols for identifying, documenting, and communicating many, typically tacit, aspects of scientific practice. Indeed, recent ethnographic works show how visions of AI in science are driving precisely this type of explication of previously tacit procedural knowledge and expertise at an unprecedented rate and scale, whether carried out by scientists themselves (in order to be able to use systems that require such context) or by institutionally hired automation engineers~\citep{nelson2025automating}. Nonetheless, work on challenges in developing effective transparency documentation for open science in general~\citep{nosek2015promoting,stodden2014implementing} and social applications of AI systems in particular~\citep{winecoff2025improving,pratt2025documenting} illustrate potential difficulties.

Institutional considerations also extend to factors moderating personalized AI's impacts. Beyond governance tools for preventing overuse, depending on the design of such systems, coordinated changes in exploratory practices may be required for better leveraging their capabilities. Our results suggest this might be particularly salient in decomposable problems: when certain AI systems become effective in autonomously dealing with or assisting in self-contained computational aspects of the scientific practice, this effectiveness would shift the bottlenecks of epistemic practice to the non-computational module. This, in turn, calls for institutional restructuring, such as patterns of re-skilling and division of labor that emphasize non-computational expertise while maintaining sufficient computational knowledge and literacy to oversee AI interaction effectively.

This consideration supports broader efforts toward designing AI systems that \textit{complement} rather than replicate human expertise~\citep{rastogi2023taxonomy,
steyvers2022bayesian}. This is particularly importance, since, in contrast to our default simulations where agents' capabilities are randomly drawn and jointly cover all problem dimensions equally well, in practice, scientific communities may exhibit systematic strengths or weaknesses. When such patterns exist, AI systems that address gaps and augment existing capabilities may provide further gains. Notably, this is in contrast to prevalent AI development practices. For example, recent analyses of clinical AI benchmarks reveal misalignment between AI capabilities and practitioner needs, with benchmarks prioritizing tasks clinicians already handle well, while neglecting those where AI could provide the greatest complementary value~\citep{blagec2023benchmark}. Our results are thus in line with calls for a shift in development priorities: rather than replicating existing human strengths, AI systems should be designed to target capability gaps to enable a productive division of labor.\footnote{Recent proposals about ``human-aware'' AI that are tuned to generate promising hypotheses that are likely to be neglected by current epistemic communities also align with this suggestion~\citep{sourati2023accelerating}.}

\subsubsection*{The Value-Laden Nature of AI Preparedness}
More broadly, our results suggest that epistemic communities cannot simply adopt AI tools and expect benefits. Rather, successful AI adoption requires prior investment in shared practices and institutional infrastructure. Context legibility and complementary expertise are such potential preconditions. Even, problem modularity, which enables portability of solutions, is rarely a fixed property of scientific problems, but rather an achievement of institutional practices around standardization, protocol development, workflow redesign, and divisions of disciplinary labor~\citep{reijula2023division,kitcher1990division,leonelli2019data,fujimura1987constructingdo,bowker2000sorting}.\footnote{Unsurprisingly, recent empirical findings show the failure of AI initiatives that ignore those prerequisite organizational factors~\citep{yee2025agentic,van2023ai}.}

But, if realizing visions of AI-driven science is contingent on such institutional investment, then the choice to pursue AI preparedness becomes a \textit{value-laden} and \textit{context-dependent} question. How we prepare and what we choose to invest in depends on balancing many competing values and making judgments about which types of questions are most pressing, given resource constraints and existing capabilities. Even assuming a common aim of epistemic progress, the way forward is not automatically clear. If in a given domain, the tasks facing the most acute resource constraints are non-computational, then, rather than depending on promised benefits of widespread AI adoption, scientific communities must seriously consider whether greater returns might not come from investing directly in those non-computational dimensions.\footnote{This is not to say that judicious use of AI, such as to address existing capability gaps, as discussed above, cannot be beneficial even in under-resourced settings. Rather, the point concerns the broader allocation of resources and attention.}

\subsection{Assumptions, Limitations, and Future Directions}
Like any other model, the framework proposed here relies on abstractions and idealizing assumptions that limit its direct applicability. Below we clarify some of these assumptions as a way to both better situate the implications of our findings, and identify directions for future research. One such assumption concerns the boundary between computational and non-computational decisions. In our model, this partition is fixed and exogenously specified. As mentioned in the previous subsection, however, in real scientific practice this boundary is neither neutral nor static. Understanding how such boundaries are constructed, and how they evolve alongside AI capabilities is therefore an important direction for future work.

Consider, moreover, two assumptions underlying the Personalized AI recommendations. First, our model treats an agent's current research practice (their decision configuration) as fully explicit and directly observable, whereas in practice such context needs to be inferred and made explicit---whether by AI systems, automation engineers, or users themselves~\citep{nelson2025automating}---from incomplete, noisy, or previously tacit information. Second, in our model, the Personalized AI perfectly predicts consequences of hypothetical incremental changes to $m_c$ in virtue of perfect access to the true payoffs of the resulting configurations, rather than relying on inferred patterns in observed configurations to predict unseen ones.\footnote{Though we also implement an imperfect and error-prone version in Section~5.} In practice, both forms of knowledge---of user context and of outcome payoffs---are inferred, partial, and error-prone, which could undermine the benefits of personalization as modeled here. 

Our model also abstracts away from factors that are necessary for understanding additional risks and negative externalities associated with different AI designs. Personalization, for example, can result in potential amplification of confirmation bias at the individual level or polarization at the collective level~\citep{kirk2024benefits}. Yet, in our model, agents are assumed to have direct access to fitness values of the landscape, and so cannot be said to possess belief-like representational states---that is, states that could \textit{misrepresent} their target. As a result, our model does not capture how the conduct of epistemic agents may be biased by their particular pre-existing beliefs. For this reason, given that our model does not capture those risks, our results should not be interpreted as a naively optimistic endorsement of personalization. 

Similarly, our model does not capture the full costs associated with misleading or useless recommendations. In the simulations, agents immediately and accurately evaluate the fitness of any proposed change, and thus quickly reject recommendations that do not improve their epistemic position. Useless recommendations, therefore, incur only an opportunity cost---agents forego a round of social learning or individual exploration. In real scientific practice, by contrast, researchers can lack the expertise or information needed to independently assess the quality of a suggested method or model, precisely in the domains where AI assistance is most attractive. In such cases, seemingly promising recommendations may have subtle or delayed adverse effects, and the harms of adopting a misguided approach---from misallocated resources and effort to retracted findings---may only become visible much later and are not easily mitigated~\citep{ehsan2022algorithmic,lacroix2021dynamics}. As a result, the opportunity costs of relying on inappropriate or context-mismatched AI suggestions are likely to be far greater in real scientific settings than our idealized model can reveal.

Another simplifying assumption of our model is that the incentives of individual agents are static and independent: an agent's payoff depends solely on the fitness of its own research practice, and not on the discoveries or timing of discoveries made by others. As a result, the model does not capture dynamics in which the value of a contribution or discovery depends on when or by whom it is made~\citep{kitcher1990division,strevens2003role}. This abstraction allows us to examine the impacts of AI adoption on communities as a whole under different structural conditions, but it can obscure important tensions between individual and collective incentives. 

Exploring such disconnects between individuals or groups and communities in light of AI adoption offers interesting directions for future research. For instance, across all AI types, we observe an inverted-U relationship between community-level performance and the rate of AI use. Future work could explore whether AI adoption creates a tragedy of the commons, where individual incentives to query AI for short-term efficiency diverge from the community's long-term need for diverse exploration. Such dynamics would constitute an additional mechanism of homogenization that our current model does not capture, and would highlight the need for governance mechanisms to realign individual benefits with collective epistemic success. 

Finally, our simulations assume uniform potential for access across the community. Real scientific environments, however, are often characterized by significant resource disparities. Future research can investigate the emergent dynamics that arise when AI tools are available only to specific subgroups. By modeling these disparities within structures where access is correlated with network position, for example, researchers can examine how AI adoption interacts with antecedent power asymmetries and other forces of homogenization. Such work could reveal whether selective AI access disrupts existing epistemic hierarchies or creates new forms of inequality.

Taken together, our analysis underscores that the epistemic impact of AI integration depends as much on the social organization of inquiry as on the design of the tools themselves. The risks we identify arise even under the aforementioned idealized and simplifying assumptions. In real scientific communities, where expertise and resources are unevenly distributed, incentives are complex, these constraints will likely be more acute. Avoiding the trap of epistemic monoculture requires shifting our focus from the passive adoption of new technologies to the active design of the institutions that wield them.

\appendix
\section{Appendix}\label{sec:appendix}
In Section 4, we introduced Non-personalized AI (NP), which recommends the computational module of the globally best-performing agent without regard for the querying agent's context. In Section 5, we introduced Single-bit Personalized AI (SBP), which identifies the single best bit to flip in the querying agent's computational module \textit{given their full decision vector}. These two designs differ along multiple dimensions simultaneously, making it important to clarify which design feature drives the observed performance differences.\footnote{We are grateful to an anonymous reviewer for emphasizing this point, and encouraging us to conduct further experiments to disentangle these dimensions.} Specifically, NP and SBP differ along at least three dimensions:

\begin{enumerate}
    \item \textbf{Scope of suggestions.} NP replaces all $M$ computational bits at once, whereas SBP modifies only a single bit.
    
    \item \textbf{Source of the candidate suggestions.} NP draws its candidate configurations from the existing community---specifically, the $n-1$ computational modules currently in use by other agents. SBP's candidates are the $M$ single-bit variants of the focal agent's current computational module. These are counterfactual configurations in the immediate neighborhood of the agent's current state, and need not correspond to any agent's existing practice.

    \item \textbf{Context-sensitivity of the evaluation procedure.} NP evaluates candidates in a context-insensitive way. In effect, NP assumes that the potential outcome of the focal agent adopting a candidate configuration is directly related to the observed fitness of that candidate's current holder, thus ignoring potential differences between the two agents' non-computational decision contexts. SBP, in contrast, evaluates candidates context-sensitively. It perfectly estimates the fitness each candidate would produce in the non-computational context of the querying agent, based on the true fitness of candidate configurations, and recommends the one with the highest fitness.\footnote{We also test an error-prone version of SBP in which the system identifies the best bit to flip with a certain accuracy but may recommend a random bit instead; see Section 5. The qualitative patterns are unchanged.}
\end{enumerate}

To further disentangle these dimensions, we introduce an additional AI design that modifies only the evaluation procedure (dimension 3) relative to NP, while holding scope and source of the candidate pool fixed.

\subsection{Personalized Chunk-Copy AI (PCC)}
Personalized Chunk-Copy AI (PCC) introduces personalization into the NP AI system, while preserving its scope and source type. Like NP, PCC copies the entire computational module ($M$ bits) from an existing agent in the community. Unlike NP, the choice of \emph{which} agent to copy from is personalized: PCC evaluates each community member's computational module in the context of the querying agent's non-computational decisions, and recommends the one that yields the highest fitness for the querying agent.

\begin{figure}[ht]
    \centering
    \begin{subfigure}[b]{0.32\textwidth}
         \centering
         \includegraphics[width=\textwidth]{Figures/NP-fitness-heatmap1000.png}
         \caption{}
         \label{fig:non-personalized-fitness-H10-appendix}
    \end{subfigure}
    \begin{subfigure}[b]{0.32\textwidth}
         \centering
         \includegraphics[width=\textwidth]{Figures/BC-fitness-heatmap1000.png}
         \caption{}
         \label{fig:BC-personalized-fitness-H10-appendix}
    \end{subfigure}
    \begin{subfigure}[b]{0.32\textwidth}
         \centering
         \includegraphics[width=\textwidth]{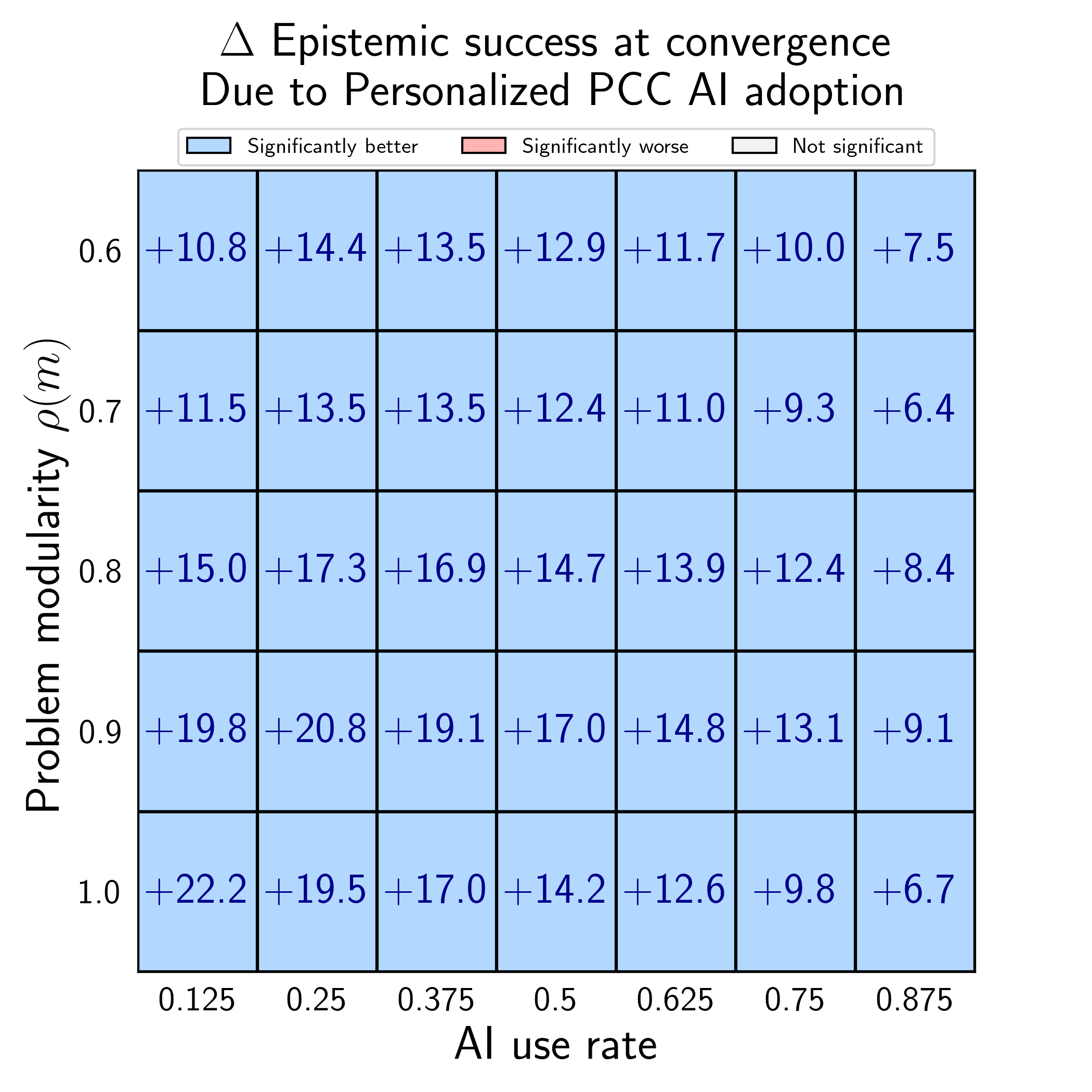}
         \caption{}
         \label{fig:PCC-personalized-diversity-H10-appendix}
    \end{subfigure}
    \caption{The differential impacts of introducing different types of AI systems into epistemic communities, expressed as percentage differences in epistemic success at convergence across problem modularity and AI use frequency, relative to communities without AI tools (averaged across 1000 simulation runs). Communities consist of $100$ agents with specialization set size $H=10$, navigating a problem with $N=20$, $K=9$, $M=10$, while engaging in social learning with $p_{social}=0.1$. Panel (A) shows the impacts of introducing \textit{non-personalized AI}. Panel (B) shows the impacts of introducing \textit{SB personalized AI}. Panel (C) shows the impacts of introducing \textit{PCC personalized AI}. Colored cells indicate statistically significant ($p < 0.05$) differences: blue for improvements, red for declines, gray for non-significant ($p >= 0.05$).}
    \label{fig:personalized-appendix}
\end{figure}

PCC thus isolates the contribution of context-sensitivity. If PCC substantially outperforms NP, this indicates that tailoring the recommendation to the querying agent's $\neg m_c$ configuration is a key driver of the performance differences documented in Sections 4 and 5, independent of the scope or candidate pool differences between NP and SBP. As Figure~\ref{fig:personalized-appendix} shows, this is indeed what we observe.  

\subsection{Why SBP Remains the Primary Variant}
While PCC is a useful diagnostic, we retain SBP as our primary personalization variant because PCC is considerably more epistemically demanding than SBP. To evaluate candidates context-sensitively, PCC must predict the fitness of $n - 1$ counterfactual configurations per query---one for every other agent's computational module combined with the querying agent's non-computational context. Importantly, each of these configurations may be radically different from the querying agent's current practice. In contrast, SBP evaluates only $M$ configurations, each differing from the agent's current practice by a single bit.

As a result, the ``perfect estimation'' assumption involved in context-sensitive evaluation implies different degrees of demandingness in the two designs. For SBP, it amounts to accurately predicting the fitness consequences of small, local changes to the agent's own practice. This is essentially the same type of iterative local search that epistemic agents perform on their own, but accelerated: instead of randomly selecting one bit to explore per round, SBP evaluates all $M$ neighbors simultaneously and recommends the best one. And it is the kind of incremental assessment that is broadly analogous to emerging AI-assisted scientific tools such as autonomous laboratories that iteratively propose and refine experimental parameters within a researcher's ongoing line of inquiry~\citep{szymanski2023autonomous}. For PCC, perfect estimation means accurately evaluating the fitness consequences of combining any community member's entire computational approach into any other querying agent's non-computational context. This is a far more demanding form of knowledge with no clear real-world analogue. 

PCC thus serves as a diagnostic tool for establishing that context-sensitivity drives the performance advantage, while SBP remains the more interpretively grounded design for the paper's central analysis.

\bibliographystyle{ACM-Reference-Format}
\bibliography{refs}

\end{document}